\documentclass[prb,aps,twocolumn,footinbib,floatfix,10pt,longbibliography,superscriptaddress]{revtex4-1}

\usepackage[english]{babel}
\usepackage{braket}
\usepackage{breakcites}
\usepackage{bm,braket}
\usepackage{float}
\usepackage{graphicx}
\usepackage[caption=false]{subfig}
\usepackage{amsfonts}
\usepackage{amsmath}
\usepackage{amssymb}
\usepackage{color}

\usepackage[utf8]{inputenc}
\usepackage[T1]{fontenc}

\newcommand {\be}{\begin{equation}}
\newcommand {\ee}{\end{equation}}

\newcommand {\bea}{\begin{eqnarray}}
\newcommand {\eea}{\end{eqnarray}}
\newcommand {\nn}{\nonumber}

\def\r1{\textbf{r}}
\def\nn{{\bf n}_d {\bf n}_d^{\dagger}}

\newcommand{\mr}{\mathbf{r}}

\newcommand{\me}{\mathbf{e}}

\usepackage{hyperref}
\usepackage[colorinlistoftodos,prependcaption]{todonotes}

\usepackage{xargs}
\newcommandx{\greencom}[2][1=]
{\todo[inline, color=green!40,#1]{#2}}
\newcommandx{\bluecom}[2][1=]
{\todo[inline, color=blue!40,#1]{#2}}
\newcommandx{\bluemargin}[2][1=]
{\todo[color=blue!40,#1]{#2}}

\graphicspath{{./figs/}}

\usepackage{letltxmacro}

\LetLtxMacro{\ORIGselectlanguage}{\selectlanguage}
\makeatletter
\DeclareRobustCommand{\selectlanguage}[1]{%
  \@ifundefined{alias@\string#1}
    {\ORIGselectlanguage{#1}}
    {\begingroup\edef\x{\endgroup
       \noexpand\ORIGselectlanguage{\@nameuse{alias@#1}}}\x}%
}
\newcommand{\definelanguagealias}[2]{%
  \@namedef{alias@#1}{#2}%
}
\makeatother

\definelanguagealias{en}{english}
\definelanguagealias{EN}{english}

\begin{document}
\sloppy
\emergencystretch 3em


\title{Theory and experiments of coherent photon coupling in semiconductor nanowire waveguides with quantum dot molecules}

\author{Chelsea Carlson}
\affiliation{Department of Physics, Queen's University, Kingston, Ontario, Canada, K7L 3N6}
\email{0clac@queensu.ca}
\author{Dan Dalacu}
\affiliation{National Research Council of Canada, Ottawa, Ontario, Canada, K1A 0R6}
\author{Chris Gustin}
\affiliation{Department of Physics, Queen's University, Kingston, Ontario, Canada, K7L 3N6}
\author{Sofiane Haffouz}
\affiliation{National Research Council of Canada, Ottawa, Ontario, Canada, K1A 0R6}
\author{Xiaohua Wu}
\affiliation{National Research Council of Canada, Ottawa, Ontario, Canada, K1A 0R6}
\author{Jean Lapointe}
\affiliation{National Research Council of Canada, Ottawa, Ontario, Canada, K1A 0R6}
\author{Robin L. Williams}
\affiliation{National Research Council of Canada, Ottawa, Ontario, Canada, K1A 0R6}
\author{Philip J. Poole}
\affiliation{National Research Council of Canada, Ottawa, Ontario, Canada, K1A 0R6}
\author{Stephen Hughes}
\affiliation{Department of Physics, Queen's University, Kingston, Ontario, Canada, K7L 3N6}


\date{\today}

\begin{abstract} 
We present a quantum optics theory, numerical calculations, and experiments on coupled quantum dots in semiconductor nanowire waveguides. We first present an analytical Green function theory to compute the emitted spectra of two coupled quantum dots, treated as point dipoles, fully accounting for retardation effects, and demonstrate the  signatures of coherent and incoherent coupling through a pronounced splitting of the uncoupled quantum dot resonances and modified spectral broadening. 
{In the weak excitation regime,} 
the classical Green functions used in models  are verified and justified through full 3D solutions of Maxwell equations
for nanowire waveguides, specifically using finite-difference time-domain techniques, showing how both waveguide modes and near-field evanescent mode coupling is important.
{The theory exploits
an  ensemble-based quantum description, 
and and an intuitive eigenmode-expansion based Maxwell theory.}
We then demonstrate how the {molecular  resonances} (in the presence of coupling) take on the form of  bright and dark (or quasi-dark) resonances, and study how these depend on the excitation  and detection conditions.
{To go beyond the weak excitation regime,}
we also introduce a quantum master equation approach
to model the nonlinear spectra from an increasing incoherent pump field, which shows the role of the pump field
on the oscillator strengths and broadening of the molecular resonances, with and without pure dephasing. 
Next, we present experimental photoluminescence spectra for spatially-separated quantum dot molecules (InAsP)  in InP nanowires, which show {clear signatures of pronounced splittings, though they also highlight additional mechanisms that are not accounted for in the dipole-dipole coupling model}.
Two different approaches are taken to control the 
spatial separation of the quantum dot molecules, and we discuss
the advantages and disadvantages of each.

\end{abstract}


\maketitle


\section{Introduction}

Quantum dots (QDs) have gained substantial interest over the years for controlling and manipulating photons and light-matter interactions. These mesoscopic semiconductor `islands'  act as  artificial atoms with bound exciton states that make them promising candidates for single photon sources \cite{somaschi_near-optimal_2016,senellart_high-performance_2017}, entangled photon pairs~\cite{jons_bright_2017}, and even photon triplets~\cite{khoshnegar_solid_2017}. 
One of the main challenges with working with single QDs, rather than ensembles, is control over the size, shape, position, and composition of the dot such that it maintains specific quantum emission characteristics~\cite{lodahl_interfacing_2015} (i.e. to couple to a cavity mode \cite{calic_deterministic_2017}). This problem is even more significant when one requires two or more QDs that can be resonantly coupled, due to their very small spectral linewidths. 

Recently, there has been much progress in bottom-up nanowire based quantum dots. In this system, the dot geometry can (in principle) be controlled with unprecedented precision\cite{Bjork_APL2002}. The nanowires are readily grown using site-selective approaches\cite{Bjork_NL2002,Dalacu_NT2009}, with each device inherently containing a single emitter thus facilitating high purity single photon emission\cite{dalacu_ultraclean_2012}. The position-control does not affect the optical quality of the emitter, with demonstrated near-lifetime-limited linewidths of 4\,$\mu$eV and post-selected Hong-Ou-Mandel visibilities exceeding 80\%\cite{Reimer_PRB2016}. The nanowire approach provides a highly symmetric system with reduced fine structure splitting\cite{Singh_PRL2009}, facilitating polarization entangled pair generation via the biexciton-exciton cascade\cite{Versteegh_NC2014,Huber_NL2014,jons_bright_2017}. In a properly designed waveguide\cite{Gregersen_OL2008,Friedler_OE2009}, the devices can be extremely bright, with efficiencies of 43\% reported\cite{Reimer_PRB2016} and a potential for higher performance through the use of a back mirror\cite{Friedler_OL2008}.

Uniquely, nanowire systems provide a controlled platform for incorporation of (nominally) perfectly aligned quantum dots optimally coupled to a mutual optical mode\cite{Borgstrom_NL2005,khoshnegar_solid_2017,haffouz_bright_2018} with a dot to dot separation controlled to the precision available to epitaxial growth techniques. Such a platform is required for enabling strong photon coupling between the dots. Coupling QDs together opens up a rich range of coherent coupling effects, such as QD entanglement~\cite{PhysRevLett.80.2245}, quantum state transfer~\cite{cirac_quantum_1997}, waveguide-mediated superradiance~\cite{Minkov2013,PhysRevA.95.053807},  and the ability to manipulate flying qubits~\cite{yao_theory_2005}.

In this work, we present a detailed theory to describe  how the spectral signatures of  QD molecules in nanowire systems can give clear signatures of photon transport in the strong QD-QD coupling regime; we then show related experiments for QD molecules in InP nanowire waveguides. 
In Sec.~\ref{sec:theory}, we first present a photon Green function (GF) theory of light propagation in nanowires, and derive an expression for the emitted spectrum in the presence of two dipoles. We also introduce analytical GFs for a homogeneous medium and
the waveguide mode of a nanowire, and show later how these can be used to construct an accurate analytical model that includes both waveguide medium transport and near field
evanescent coupling (e.g., F\"orster coupling, and other
photon coupling effects from near field interactions). We present both classical and quantum expressions for the fields, and use the latter to derive the spectrum for excited quantum dot states in vacuum. 
{The GF approach is restricted to weak excitation and equivalent results can be derived classically. To go beyond this approach, and
to investigate the nonlinear regime of strong pumping in the emission spectrum, as well as linewidth broadening through pure dephasing, we also employ a Markovian master equation approach for coupled QDs} in Sect.~\ref{sec:me} and assess its validity through comparison with the GF approach.
In Sec.~\ref{sec:results}, we  show  the key features of the medium GFs, both analytically and numerically, using full 3D finite-difference time-domain (FDTD) calculations, and develop an intuitive
analytical model for the QDs in a nanowire waveguide. Then we use these models to compute the emitted spectrum, investigating various features such as detector position and the sensitivity to the initial excitation, and QD separation. We highlight various spectral features, and discuss the emergence of dark and bright states,  as well as superradiance and sub-radiance features. Through the nonlinear excitation master equation solution, we also highlight additional spectral features, including a reversal of the relative spectral weights of peaks in the emission spectrum, that arise from two-exciton states, which can achieve significant population under strong pumping conditions.
In Sec.~\ref{sec:exp}, we show experimental results for
QD molecules in InP nanowires, where we see a pronounced spectral splitting between excitons that increases for smaller dot-dot separations, though there are clearly other mechanisms that are not captured within our quantum dipole
model for the quantum dot excitons.
Finally, in Sec.~\ref{sec:con}, we present our conclusions. {In addition, 
we include three appendices.
Appendix~\ref{app:1}
discusses a classical oscillator approaches to model 
finite size dipoles in FDTD}, 
Appendix~\ref{appendix:ME} presents some technical details about the master equation solution for the nonlinear incoherent spectrum,
and Appendix~\ref{appendix:data}
presents various experimental spectra for various sets
of QD nanowires.

\section{Green Function Theory} 
\label{sec:theory}

\subsection{Photon Green function for a lossless  waveguide mode}
Light propagation through an arbitrary dielectric medium can be described in terms of the mode solutions to the Helmholtz equation: 
\begin{equation}
\nabla \times \nabla \times \mathbf{f}_{\lambda}({\bf r}) - \frac{\omega_{\lambda}^2}{c^2} \epsilon(\mathbf{r})\mathbf{f}_\lambda(\mathbf{r})=0,
\label{eq:helmholtz}
\end{equation}   
where $\epsilon({\bf r})$ describes the relative permittivity of the structure and $\mathbf{f}_\lambda(\mathbf{r})$ are generalized field modes with harmonic $e^{-i\omega t}$ time dependence.  The electric-field GF \cite{Minkov2013}, which describes the field response at  $\mathbf{r}$ to a point source at $\mathbf{r'}$ is defined through
\begin{equation}
\left [\nabla \times \nabla \times - \frac{\omega^2}{c^2} \epsilon(\mathbf{r})\\ \right ]\mathbf{G}(\mathbf{r},\mathbf{r}',\omega) =\frac{\omega^2}{c^2}{\mathbf{1}} \delta(\mathbf{r}-\mathbf{r'}),
\label{eq:greens}
\end{equation}
where ${\mathbf G}_{i,j}$ is a second rank tensor and $\mathbf{1}$ is the unit dyad; elements [$i, j$] correspond to the response in direction $i$ at $\mathbf{r}$ from the $j$th component of the source at $\mathbf{r'}$.  When the GF is known, the field response to an arbitrary polarization dipole source  $\mathbf{P}(\mathbf{r}, \omega)$ can be found from
\begin{equation}
\begin{aligned}
\mathbf{E}(\mathbf{r}, \omega)&=\mathbf{E}^{\rm h}(\mathbf{r}, \omega)
+\frac{1}{\epsilon_o}\int_{V'} \mathbf{G}(\mathbf{r}, \mathbf{r'}; \omega)\cdot \mathbf{P}(\mathbf{r'}, \omega)d\mathbf{r'},
\label{eq:gdef}
\end{aligned} 
\end{equation}
in which $\mathbf{E}^{\rm h}$ is the homogeneous  field solution in the absence of the polarization source.  The eigenmodes of Eq.~\eqref{eq:helmholtz}, $\mathbf{f}_\lambda(\mathbf{r})$,  form an orthonormal  and  complete set,
so that 
$\int_V \epsilon(\mathbf{r})\mathbf{f}_\lambda(\mathbf{r})\cdot \mathbf{f}_{\lambda'}^*(\mathbf{r})d\mathbf{r} =\delta_{\lambda, \lambda'}$
and 
$\sum_ {\lambda}\epsilon(\mathbf{r})\mathbf{f}_{\lambda}(\mathbf{r}) \mathbf{f}_{\lambda}^*(\mathbf{r'}) ={\mathbf{1}} \delta(\mathbf{r}-\mathbf{r'})$\cite{Sakoda2005}. 

The waveguides of interest here are photonic nanowires,
which have discrete translational symmetry in their in-plane dielectric structure, supporting lossless waveguide modes
$\mathbf{f}_{k_\omega}(\mathbf{r})=\sqrt{\frac{1}{L}}\mathbf{e}_{k_\omega}(\bm{\rho})e^{i k_\omega z }$, where $\mathbf{e}_{ k_\omega}(\bm{\rho})$ is the mode solution, normalized according to $\int_{A_{\rm w}} \epsilon(\bm{\rho})|\mathbf{e}_{ k_\omega}(\bm{\rho})|^2d\bm{\rho}=1$, where $A_{\rm w}$ is the spatial area, $k_{\omega} = \omega n_{\omega}/c$, $n_{\omega}$ is the effective index, and $L$ is the length of the structure. One can then obtain the waveguide mode  GF analytically as~\cite{Rao2007theory}
\begin{align}
\mathbf{G}_{\rm WG}(\mathbf{r}, \mathbf{r'}, \omega) &= 
\frac{i \omega}{2 v_g}\Big[\Theta(z-z')\mathbf{e}_{k_\omega}(\bm{\rho})\mathbf{e}^*_{ k_\omega}(\bm{\rho}')e^{i k_\omega (z-z') } 
\nonumber \\
+&\Theta(z'-z)\mathbf{e}^*_{k_\omega}(\bm{\rho})\mathbf{e}_{ k_\omega}(\bm{\rho}')e^{i k_\omega (z'-z) } \Big],
\label{eq:Gwg}
\end{align}
where the terms preceded by Heaviside functions correspond to forward and backwards propagating modes, respectively, and $v_g=|v_g(\omega)|$ is the group velocity at the frequency on interest. Since the modes are translationally invariant in $z$, 
$\mathbf{e}_{k_\omega}({\bf r}_d)=\mathbf{e}_{k_\omega}(\bm{\rho}_d)$,
where ${\bf r}_d$ is the quantum dot (QD) position (and we assume the QD is at the center of the wire axis, $x=y=0$). For simplicity, we will also introduce
the peak field position, which is maximally coupled to the waveguide mode, both in terms of position and polarization, such that
$\mathbf{e}_{k_\omega}({\bf r}_0)=\mathbf{e}_{k_\omega}(\bm{\rho}_0)$,
with $|\mathbf{e}_{k_\omega}({\bf r}_0)|^2={1}/{(A_{\rm eff}\epsilon_{\rm B})}$, with $A_{\rm eff}$ the effective mode area, and $\epsilon_{\rm B}=n_{\rm B}^2$ the bulk background dielectric constant of the photonic wire.

\subsection{Homogeneous medium Green function}

In the near field, the homogeneous medium GF contribution
can be the dominant coupling mechanism in various
inhomogeneous dielectric systems, so we discuss the general properties here and confirm when this is a good approximation in the results section below.
Thehomogeneous medium GF is
\begin{equation}
\begin{aligned}
{{\bf G}_{\rm hom}}({\bf r},{\bf r}';\omega) = &\frac{k_0^2e^{ik_{\rm B}R}}{4\pi R}\bigg[\bigg( 1 + \frac{i}{k_{\rm B}R}-\frac{1}{(k_{\rm B}R)^2}\bigg){\mathbf{1}} + \\
&
\ \
\bigg(\frac{3}{(k_{\rm B}R)^2} - \frac{3i}{k_{\rm B}R} - 1\bigg) \frac{\mathbf{R}\otimes\mathbf{R}}{R^2}\bigg],
	\label{eq:GFanalytic}
\end{aligned}
\end{equation}
where ${\bf R}={\bf r}-{\bf r}'$, $k_{\rm 0}=\omega/c$, $k_{\rm B}=n_{\rm B}\omega/c$ and ${\bf 1}$ is the unit dyadic.
We will show later that the homogeneous GF is one of the dominant contributions to the total GF for dot-dot coupling in the near field of a nanowire. However, in general one also needs the waveguide mode GF as well, e.g., to describe photons propagating along the wire, and to satisfy the optical theorem.

For two point dipoles with $x$ and/or $y$ dipole moments, but separated in $\mathbf{\hat{z}}$ (see Fig.~\ref{fig:SpectraSetup}), we can write the near field ($k_BR\ll 1$) GF as
\begin{equation}
G_{\rm{hom}}|_{xx,yy} = \frac{-1}{4\pi n_{\rm B}^2 R^3} + i \frac{n_{\rm B}}{6\pi}\bigg(\frac{\omega}{c}\bigg)^3, 
\label{eq:ghom}
\end{equation}
neglecting the intermediate and far field contributions ($1/R^2$ and $1/R$ terms, respectively). 
This simplification if often used in the literature
and makes a clear connection to typical dipole-dipole
coupling models in the quasistatic limit.
Note that since we treat perfect point dipoles, we do not allow
for any effects beyond the dipole approximation, and the possibility of polarization mixing between the
$x$ and $y$ dipole moments.
While it would be interesting to explore such effects, we will neglect them in this work.

\subsection{Coupling classical fields to a single quantum dot exciton or two level atom}

It is useful to first
 consider an embedded  single QD  or two level atom treated
at the level of a {\it classical} polarization dipole. We assume, for now, that
the polarizability of the QD exciton, with resonance energy $\omega_0$, is described through the polarizability tensor
\be
{\bm \alpha}_0 = \alpha_0  \nn  ,
\ee
where ${\bf n}_d = a \hat{\bf x} + b  e^{i\phi} \hat{\bf y}$ is a unit vector with some arbitrary in-plane polarization direction, with $a^2+b^2=1$,
$\phi$ some arbitrary phase,
and
\be
\alpha_0 = \frac{2 \omega_0 d_0^2/\epsilon_0\hbar}{\omega_0^2-\omega^2},
\label{alpha}
\ee
is the ``bare polarizability'' volume, i.e., it does not include radiative coupling effects to the environment. For simplicity, we will also neglect nonradiative
decay processes (unless stated otherwise)\ and assume that radiative coupling is the dominant decay mechanism, though this can easily be added into the above response function prior to adding in  radiative coupling. The total electric field in the wire waveguide can now be written as
\begin{align}
\mathbf{E}(\mathbf{r}, \omega)&=\mathbf{E}^{\rm h}(\mathbf{r}, \omega)
+ \mathbf{G}(\mathbf{r}, \mathbf{r}_{d}; \omega)\cdot 
{\bm \alpha}_0(\omega) \cdot {\bf E}({\bf r}_d,\omega),
\label{eq:eprop1}
\end{align} 
where ${\bf G}$ has units of inverse volume
and ${\bm \alpha}_0$ has units of volume. Since the QD is polarized in the plane of the wire (i.e., in $\hat {\bf x}$ and $\hat {\bf y}$) and the lowest-order propagating mode for the wire waveguide is $HE_{11}$ mode (also polarized in the plane) ~\cite{Bulgarini2014}, we only need to consider the GF as a two by two matrix, e.g., in a Cartesian coordinate system,
\be
{\bf G} = \begin{pmatrix}G_{xx} & G_{xy} \\
	G_{yx} & G_{yy} \\
\end{pmatrix}.
\ee
Note that we can choose any basis we like, as long as it is complete,
so we do not need to choose a linearly polarized basis. This can important for more general QD coupling, such as
with chiral networks~\cite{young_polarization_2015,cirac_quantum_1997}.

\subsection{General quantum theory and emitted spectrum}

In this subsection, we introduce a medium-dependent quantum optics approach which closely follows the formalism of  
Refs.~\onlinecite{Yao2009,Kristensen2011}, for calculating the emission spectrum at some detection point, $\mathbf{r}_D$, for an arbitrary photonic structure with two embedded QDs, treated as point dipoles, at positions $\mathbf{r}_1$ and $\mathbf{r}_2$. This approach is particularly useful 
for highlighting the underlying physics of photon transport
in terms of classical response functions. It is also valid for 
arbitrary media, though is restricted in general to computing the linear spectra.
{Thus although we use a quantum theory below, which helps to highlight the underlying physics, all the final  equations in this subsection could be equivalently derived classically, but with a different interpretation.}
In the subsequent section, we also discuss an alternative master equation approach, which can  include nonlinear interactions.

Starting from a multipolar Hamiltonian in the dipole approximation
\cite{wubs_multiple-scattering_2004,hughes_theory_2009,Minkov2013}, the Hamiltonian of the general medium and QD dipoles is 
\begin{align}
H = &\sum_\lambda\hbar\omega_\lambda \hat a_\lambda^\dagger \hat a_\lambda + \sum_n\hbar\omega_n \sigma_n^+ \sigma^-_n  \nonumber \\
&- i\hbar\sum_{\lambda,n}\left(\sigma^+_n + \sigma^-_n\right)\left(g_{n,\lambda}\hat a_\lambda - g_{n,\lambda}^* \hat a_\lambda^\dagger \right),
\label{ham}
\end{align}
where the photon terms $\omega_\lambda$, $\hat a_\lambda^\dagger$ and $\hat a_\lambda$ are the angular frequency, creation and annihilation operators, respectively, of a photon in mode $\lambda$, and
 operators satisfy bosonic commutation relations, e.g.,
$[\hat a_\lambda, \hat a_{\lambda'}^\dagger] = \delta_{\lambda,\lambda'}$; Similarly, the dipole terms $\omega_n$, $\sigma^+_n$ and $\sigma^-_n$ denote angular frequency, creation and annihilation operators, respectively, of an electron-hole pair (an exciton) in the $n$'th QD (two dots in the case of a molecule, $n=1$ or $n=2$), and these
operators satisfy Fermion anticommutation relations, i.e.,
$\{\sigma^+_n,\sigma^-_n\} = 1$.
The light-matter coupling strength is given by
\begin{align}
g_{n,\lambda} = \sqrt{\frac{\omega_\lambda}{2\hbar\epsilon_0}}\mathbf{d}_n\cdot\mathbf{f}_\lambda(\mr_n),
\end{align}
where $\mathbf{d}_n=d_n\me_n$ is the dipole moment of the $n$'th QD of magnitude $d_n$ and orientation $\me_n$, and
$\mathbf{f}_\lambda(\mr_n)$ is the normalized mode.
Next one can derive the Heisenberg equations of motion for the photon and exciton creation and annihilation operators~\cite{wubs_multiple-scattering_2004,hughes_theory_2009}, which are then solved in the frequency domain to yield an exact expression for the operators
in the limit of weak excitation, i.e., with at most one quantum excitation in the system. 
For example, for one QD dipole
at ${\bf r}_d$, the general expression for the electric-field operator takes the form,
\be
 \mathbf{\hat E}(\mathbf{r},\omega)=\mathbf{\hat E}^{\rm h}(\mathbf{r},\omega)
+ \mathbf{G}(\mathbf{r}, \mathbf{r}_{d};\omega)\cdot 
{\bm \alpha}_n \cdot {\bf \hat E}({\bf r}_d,\omega),
\label{eq:eprop1b}
\ee
in an almost identical form to 
Eq.~(\ref{eq:eprop1}), but the quantum form can correctly describe spontaneous emission processes from vacuum fluctuations, using
an excited QD as the quantum mechanical source.
Of course the fields are now field operators~\cite{Yao2009},
and this expression also includes {\it free fields}.

Next,  let us consider an in-plane linearly polarized dipole (e.g., ${\bf d}=d \hat {\bf x}$, or ${\bf d} =d \hat {\bf y})$ in an arbitrary environment (i.e. waveguide, homogeneous, cavity, etc.). By exploiting the Dyson equation,
${\bf G}^{(1)} = {\bf G}^{} +{\bf G}^{} \cdot {\bm \alpha}_0 
\cdot {\bf G}^{(1)}$, where the `(1)' superscript denotes the GF with the
addition of one QD in the medium,
we first rewrite Eq.~(\ref{eq:eprop1b}) as ($\omega$ is implicit from now on, unless stated otherwise)
\be
 \mathbf{\hat E}(\mathbf{r})=\mathbf{\hat E}^{\rm h}(\mathbf{r})
+ \mathbf{G}^{(1)}(\mathbf{r}, \mathbf{r}_{d})\cdot 
{\bm \alpha}_n \cdot {\bf \hat E}^{\rm h}({\bf r}_d),
\label{eq:eprop2}
\ee
where the re-normalized one-dot GF is written as 
%
\begin{equation}
 \mathbf{G}^{(1)}(\mathbf{r},\mathbf{r}_n) = \frac{\mathbf{G}(\mathbf{r},\mathbf{r}_n)}{1 - {\alpha}_0 \mathbf{n}_n^\dagger\cdot\mathbf{G}(\mathbf{r}_n,\mathbf{r}_n)\cdot\mathbf{n}_n}.
\label{renorm0}
\end{equation}
Since we will use this expression at different discrete spatial points, ${\bf r}_m$, then we also define
 \begin{equation}
 \mathbf{G}^{(1)}_{m,n} \equiv \mathbf{G}^{(1)}(\mathbf{r}_m,\mathbf{r}_n) = \frac{\mathbf{G}(\mathbf{r}_m,\mathbf{r}_n)}{1 - {\alpha}_0 \mathbf{n}_n^\dagger\cdot\mathbf{G}(\mathbf{r}_n,\mathbf{r}_n)\cdot\mathbf{n}_n}.
 \label{renorm}
 \end{equation}

Note that due to the divergence of the real part of the background homogeneous GF, when $\mathbf{r}_m = \mathbf{r}_n$, we only use the imaginary part of the background homogeneous GF (i.e. $\mathbf{G}_{\rm B}(\mathbf{r}_n,\mathbf{r}_n) \rightarrow i\,\rm{Im}\{\mathbf{G}_{\rm B} (\mathbf{r}_n,\mathbf{r}_n)\}$). 
In general the GF will of course also  contain contributions from the waveguide mode, as well as contributions for radiation modes above the light line. Directly from this renormalized GF, we can rearrange to arrive at the radiative decay rate of one QD for a given dipole moment and background dielectric constant, 
\begin{equation}\label{eq:G11}
    \Gamma_{1,1}(\omega) = \frac{2d^2}{\hbar\epsilon_0}{\rm Im}[{\bf e}_1 \cdot {\bf G}(\mathbf{r}_1,\mathbf{r}_1;\omega)  \cdot {\bf e}_1].
\end{equation}

\begin{figure}[htp]
    \centering
    \includegraphics[width=0.8\columnwidth]{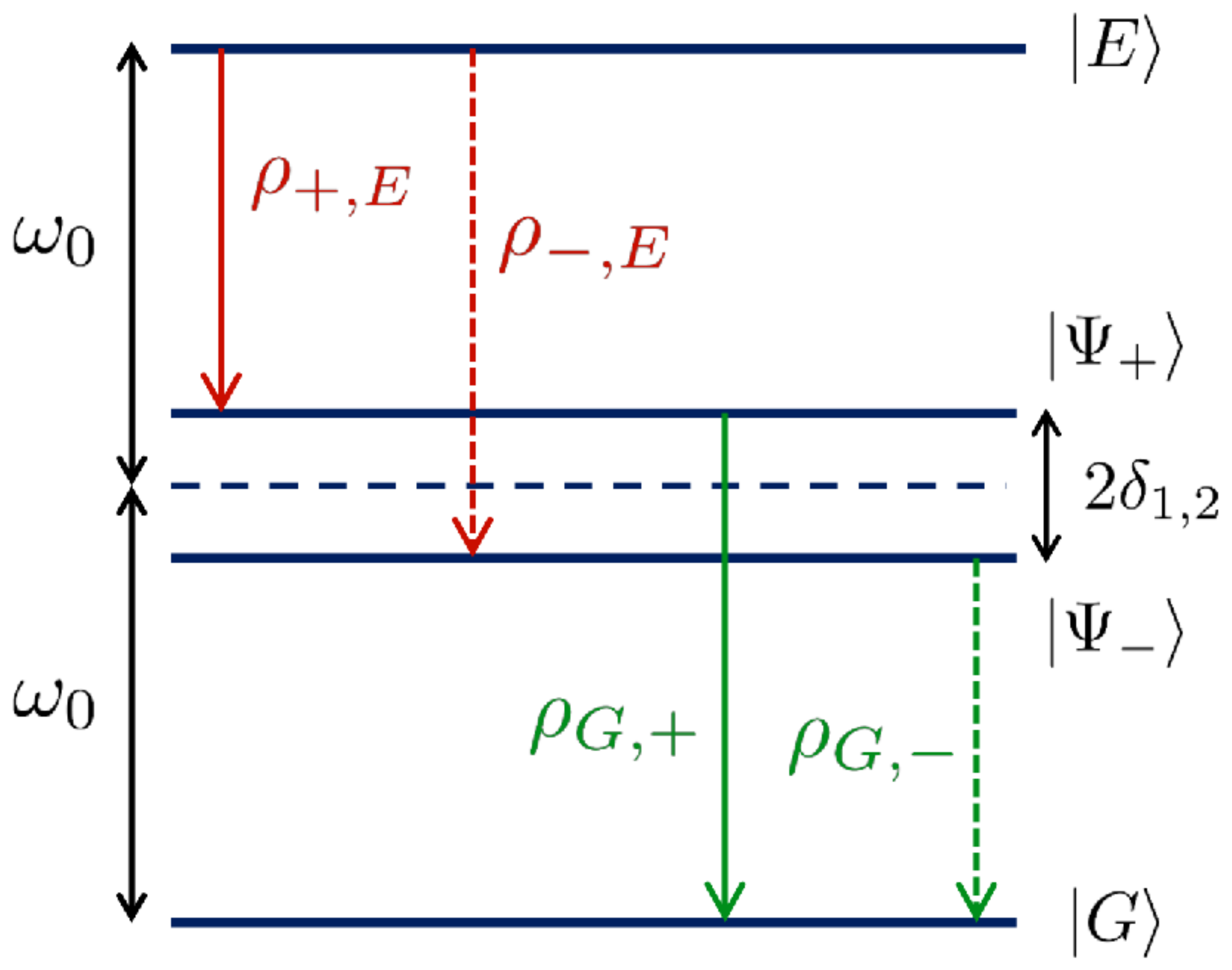}
    \caption{Energy level diagram of the coupled QD system to help formulate a quantum master equation description. In the regime of $\delta_{1,2} \gg \{\Gamma_{1,1},\Gamma',\Gamma_{1,2},\Gamma_{\rm{inc}}\}$, the emitted spectrum is a sum of Lorentzian peaks corresponding to each of the optically allowed transitions. The transitions in red involve the two-exciton $\ket{E}$ state (associated with the two excitons being excited from each QD) and thus are missed by the weak excitation approximation, but are typically important unless only one QD is pumped, or $\Gamma_{\rm{inc}} \ll \{\Gamma_{1,1},\Gamma_{1,1}-\Gamma_{1,2}\}$. The dotted transitions involving the $\ket{\Psi_-}$ state tend to interfere destructively and have smaller decay rates, and thus are typically less dominant in the emitted spectrum. We neglect the possibility of biexcitons from each individual QD.}
    \label{fig:Level_diagram}
\end{figure}

The generalization to include more than one QD is straightforward, and similar to how one solves a classical Dyson equation with multiple dipoles. In the present case,
the GF for two QD scatterers at $\mathbf{r}_1$ and $\mathbf{r}_2$, as seen by a detector at $\mathbf{r_D}$, is given by:
\begin{equation}
\mathbf{G}^{(2)}(\mathbf{r}_D,\mathbf{r}_2) = \frac{\mathbf{G}^{(1)}(\mathbf{r}_D,\mathbf{r}_2) + \mathbf{G}^{(1)}(\mathbf{r}_D,\mathbf{r}_1) 
\cdot {\bm \alpha}_1 \cdot {G}^{(1)}_{1,2}}{1 - {G}^{(1)}_{2,1}\mathbf{\alpha}_1{G}^{(1)}_{1,2}\mathbf{\alpha}_2},
\label{G2}
\end{equation}
 where ${G}^{(1)}_{i,j} = \mathbf{n}_i^\dagger\cdot\mathbf{G}^{(1)}_{i,j}\cdot\mathbf{n}_j$. A similar expression can be derived for $\mathbf{G}^{(2)}(\mathbf{r}_D,\mathbf{r}_1)$. These equations fully solve the scattering problem without any rotating-wave or  Markov approximations. As before, with some rearranging of the 2-dot renormalized GF we may arrive at expressions for the incoherent and coherent photon coupling (alternatively, the real and virtual photon transfer) rates between the two dipoles,
 \begin{equation}\label{eq:G12}
    \Gamma_{1,2}(\omega) = \frac{2d_1d_2}{\hbar\epsilon_0}{\rm Im}[{\bf e}_1 \cdot {\bf G}(\mathbf{r}_1,\mathbf{r}_2;\omega)  \cdot {\bf e}_2],
\end{equation}
\noindent and 
\begin{equation}
    \delta_{1,2}(\omega) = \frac{-d_1d_2}{\hbar\epsilon_0}{\rm Re}[\mathbf{e}_1\cdot\mathbf{G}(\mathbf{r}_1, \mathbf{r}_2,\omega)\cdot\mathbf{e}_2], 
    \label{eq:delta}
\end{equation}
 respectively, which are explicitly functions of frequency.
 The 
former modifies the broadening of the spectral resonances, while the latter {is related to} the splitting{; the  spectral splitting of light emission and detection (as would be measured) is given precisely by $2\delta_{1,2}$ (see Fig.~\ref{fig:Level_diagram})}. 

Next, the spectral response at position $\mathbf{r}_D$ can be calculated as a sum of the electric field response from the two dots, given the initial conditions,
through $\langle\sigma_{n,n}(t=0)\rangle$.
These are density matrix elements that relate to the initial populations
(on-diagonal elements)
or coherences in the system (off-diagonal elements).
This expression applies also for the field in vacuum case (i.e., no external pump field):
\begin{equation}
S(\mathbf{r}_D,\omega) \equiv \langle \mathbf{\hat E}^\dagger(\mathbf{r}_D,\omega)\mathbf{\hat E}(\mathbf{r}_D,\omega)\rangle,
\label{S1}
\end{equation}
where the electric field, $\mathbf{E}$, is given by,
\begin{equation}
\mathbf{\hat E}(\mathbf{r}_D,\omega) = \sum_{n=1,2}\mathbf{G}^{(2)}(\mathbf{r}_D,\mathbf{r}_n)\cdot \mathbf {p}_n,
\end{equation}
with the quantum oporator dipole source term,
\begin{equation}
\mathbf{p}_n \equiv \frac{i {\mathbf{d}}_n}{\epsilon_0}\bigg[ \frac{\sigma_n^-(t=0)}{\omega - \omega_n} + \frac{\sigma_n^+(t=0)}{\omega + \omega_n} \bigg]. 
\end{equation}
Thus, if the initial excitation is in the
QD dipoles, the spectrum is obtained from
\begin{align}
&S(\mathbf{r}_D,\omega)= \sum\limits_{n,m=1,2}\frac{\braket{\sigma^+_n(t=0)\sigma^-_m(t=0)}}{(\omega - \omega_n^*)(\omega - \omega_m)} \nonumber \\
&\times\frac{[\mathbf{G}^{(2)}(\mathbf{r}_D,\mathbf{r}_n;\omega)\cdot\mathbf{d}_n]^{\dagger}[\mathbf{G}^{(2)}(\mathbf{r}_D,\mathbf{r}_m;\omega)\cdot\mathbf{d}_m]}{\epsilon_0^2},
\label{eq:spectrum}
\end{align}
where, in general, the spectrum will depend upon the excitation conditions. For example, it is known that with dipole-dipole coupling effects, then the two shared excitons can take on the form of a symmetric ($\ket{\Psi_+}=1/\sqrt{2}(\ket{1}_1\ket{0}_2+
\ket{0}_1\ket{1}_2)$)  and antisymmetric state ($\ket{\Psi_-}=1/\sqrt{2}(\ket{1}_1\ket{0}_2-
\ket{0}_1\ket{1}_2$). With no additional symmetry breaking, e.g., through the propagation from the dots to the detector, the former state is super-radiant ({decaying faster than the uncoupled
dot}), and the latter is sub-radiant ({decaying slower than the uncoupled
dot}), and we thus expect an optically bright and dark resonance~\cite{tanas_entangling_2004,agarwal_quantum_2012,Thomas2002,Yao2009bell}. For later use,
we also define the states where only QD
$1$ or $2$ are excited:
$\ket{\Psi_1}= \ket{1}_1\ket{0}_2$
and
$\ket{\Psi_2}= \ket{0}_1\ket{1}_2$,
which can be realized by 
incoherent {excitation}. 
The initial electric field is taken to be in vacuum. Since the GF theory assumes weak excitation, we are considering at most one quantum excitation in the total system (weak excitation approximation)~\cite{Yao2009bell}.
In the table below, we 
summarize the various initial conditions for the QD pair
that we will consider:\\
\begin{table}[ht]
    \centering
    \begin{tabular}{|c|c|}
    \hline
       $\ket{\Psi(t=0)}$  & $\ket{\psi(t=0)}_{\rm QD}$  
       \\ \hline
         $\ket{\Psi_1}$ & $\ket{1}_1\ket{0}_2$ 
         \\
         $\ket{\Psi_2}$ & $\ket{0}_1\ket{1}_2$ \\
         $\ket{\Psi_{\pm}}$ & $\frac{1}{\sqrt{2}}(\ket{1}_1\ket{0}_2\pm\ket{0}_1\ket{1}_2)$ \\
    \hline
    \end{tabular}
    \caption{Summary of initial conditions for Eq.~\ref{eq:spectrum}.}
    \label{Tab:InitialConditions}
\end{table}


Equation~(\ref{eq:spectrum}) can be computed using analytical expressions or numerical calculations for the photon GF. However, clearly one would rather use purely analytic expressions, that we will develop below. In particular we will show how the real and imaginary parts of the medium GF can be approximated as 
\begin{align}
{\rm Re}[{\bf G}({\bf r}_1,{\bf r}_2)] &\approx {\rm Re}[{\bf G}_{\rm hom}({\bf r}_1,{\bf r}_2)] +{\rm Re}[{\bf G}_{\rm WG}({\bf r}_1,{\bf r}_2)], \nonumber \\
{\rm Im}[{\bf G}({\bf r}_1,{\bf r}_2)] &\approx {\rm Im}[{\bf G}_{\rm WG}({\bf r}_1,{\bf r}_2)], \nonumber \\ 
{\rm Im}[{\bf G}({\bf r}_n,{\bf r}_n)] &\approx (1-\beta^{\rm WG}){\rm Im}[{\bf G}_{\rm hom}({\bf r}_n,{\bf r}_n)] 
\nonumber \\
&+ 
{\rm Im}[{\bf G}_{\rm WG}({\bf r}_n,{\bf r}_n)],\label{GFs_an}
\end{align}
where $\beta^{\rm WG}$ is the waveguide beta factor, i.e., the probability that 
a photon will be emitted into the waveguide mode of the wire, discussed in more detail in the next section.

\section{Master Equation Approach and Nonlinear effects from incoherent pumping}\label{sec:me}

In this section, we introduce a master equation approach to
explore the emission spectra of incoherently excited QDs at excitation powers beyond the linear excitation regime studied in the rest of this work;
for increasing pump levels, even if below powers where the
individual biexcitons appear, the finite population of the exciton states may 
become significant enough to warrant a fully quantum mechanical treatment of the excitation dynamics beyond a weak excitation. Furthermore, linewidth broadening mechanisms (i.e., pure dephasing) beyond spontaneous emission may also be of interest. We restrict this section to the analysis of two dots with identical parameters, although we note the qualitative features (e.g., two peaks in the emission spectrum) remain in the nonlinear regime even with a detuning between the two uncoupled QDs. To study this nonlinear excitation regime, we employ a master equation for the density operator $\rho$ derived for two coupled QDs in a structured photonic reservoir.  Following Ref.~\onlinecite{Angelatos2015}, we obtain:
\begin{align}\label{master_eq}
\frac{\rm{d}\rho}{\rm{dt}} =& -\frac{i}{\hbar}[H,\rho] + \frac{\Gamma_{1,1}}{2}(\mathcal{L}[\sigma_1^-]\rho + \mathcal{L}[\sigma_2^-]\rho) \nonumber \\ &+ \frac{\Gamma_{1,2}}{2}(\mathcal{L}[\sigma_1^-,\sigma_2^+]\rho + \text{H.c.})  \nonumber \\ & + \frac{\Gamma'}{2}(\mathcal{L}[\sigma_1^+\sigma_1^-]\rho + \mathcal{L}[\sigma_2^+\sigma_2^-]\rho) \nonumber \\ & +\frac{\Gamma_{\rm{inc}}}{2}(\mathcal{L}[\sigma_1^+]\rho + \mathcal{L}[\sigma_2^+]\rho),
\end{align}
where $\mathcal{L}[A,B]\rho = 2A\rho B - BA\rho - \rho BA$, and $\mathcal{L}[A]\rho = \mathcal{L}[A,A^{\dagger}]\rho$ is the Lindblad superoperator. We have derived this equation in a frame rotating at the bare exciton frequency $\omega_0$. 
The term  $\Gamma_{1,1} = \Gamma_{1,1}(\omega_0)$ is the  spontaneous emission rate from Eq.~\eqref{eq:G11} evaluated at the bare exciton frequency as a consequence of the Markov approximation made in the master equation derivation, here equal for both QDs at positions $\mathbf{r}_1$ and $\mathbf{r}_2$. 

Similarly, with regard to the QD coupling terms, the incoherent photon transfer rate is  $\Gamma_{1,2} = \Gamma_{1,2}(\omega_0)$ from Eq.~\eqref{eq:G12}, while $\Gamma'$ and $\Gamma_{\rm{inc}}$ are phenomenologically inserted pure dephasing and incoherent pump rates, respectively. The Hamiltonian $H = \hbar\delta_{1,2}(\sigma^+_2\sigma^-_1 + \sigma^+_1\sigma^-_2)$ arises from the dipole-dipole coupling, where the coherent photon transfer rate is $\delta_{1,2}=\delta_{1,2}(\omega_0)$ from Eq.~\eqref{eq:delta}. This Hamiltonian has eigenstates $\ket{E} = \ket{1}_1\ket{1}_2$, $\ket{G} = \ket{0}_1\ket{0}_2$,  and $\ket{\Psi_{\pm}} = \frac{1}{\sqrt{2}}(\ket{1}_1\ket{0}_2 \pm \ket{0}_1\ket{1}_2)$. In the linear excitation regime (with the weak excitation approximation), the transitions $\ket{\Psi_+} \rightarrow \ket{G}$ and $\ket{\Psi_-} \rightarrow \ket{G}$ create the PL spectrum, where the second transition is rendered dark by destructive interference, barring any symmetry breaking in the propagation of the spectrum to the detector. Here, however, we pump both excitons, which introduces the possibility of exciting the $\ket{E}$ state, adding new transitions (though with the same frequency peaks as the linear spectrum) into the nonlinear emission spectrum beyond the linear GF approach. 

Additionally, for strongly coupled QDs ($\Gamma_{1,2} \approx \Gamma_{1,1}$), the spectrum can become nonlinear, even for very small pump powers -- specifically, the criterion for linearity is $\Gamma_{\rm{inc}} \ll \{\Gamma_{1,1},\Gamma_{1,1}-\Gamma_{1,2}\}$. To study the nonlinear effects, we can calculate the emission spectrum~\cite{Angelatos2015}:
 \begin{equation}
S(\omega) = \sum_{n,n'}\text{Re}\Big\{g_{n,n'}(\omega)S_{n,n'}^0(\omega)\Big\},
\end{equation}
for $n,n'=1,2$, where $g_{n,n'}(\omega) = \frac{1}{\epsilon_0^2}\mathbf{d}_n\cdot\mathbf{G}^*(\mathbf{r}_n,\mathbf{r}_D;\omega)\cdot \mathbf{G}(\mathbf{r}_D,\mathbf{r}_{n'};\omega)\cdot \mathbf{d}_{n'}$ is related to the propagation of the emitted fields to the detector at position $\mathbf{r}_D$, and
\begin{equation}
S_{n,n'}^0(\omega) = \lim_{t\rightarrow \infty}\Bigg[\int_0^{\infty}d\tau \langle \sigma_{n}^+(t+\tau)\sigma_{n'}^-(t)\rangle e^{-i(\omega-\omega_0)\tau}\Bigg],
\end{equation}
where for identical emitters, $S_{1,1}^0(\omega) = S^0_{2,2}(\omega)$, and $S_{1,2}^0(\omega) = (S^0_{2,1}(\omega))^*$. Using the quantum regression theorem~\cite{Carmichael1999}, one can exactly solve the optical Bloch equations in the basis of the system eigenstates, and decompose these spectral functions into linear combinations of Fourier-transformed density matrix elements corresponding to optical transitions between the system states (see Appendix~\ref{appendix:ME}):
\begin{equation}S_{1,1}^0(\omega) = \rho_{+,E}(\omega) - \rho_{-,E}(\omega) + \rho_{G,+}(\omega) + \rho_{G,-}(\omega),
\end{equation}
and
\begin{equation}S_{2,1}^0(\omega) = \rho_{+,E}(\omega) + \rho_{-,E}(\omega) + \rho_{G,+}(\omega) - \rho_{G,-}(\omega).
\end{equation}

Clearly, if $g_{n,n'}(\omega)$ vary little from each other over the frequency range of interest, terms corresponding to $\ket{\Psi_-}$ transitions will interfere destructively in the emitted spectrum. In the general case, these spectral transitions for the nonlinear spectrum can be found analytically:
\begin{equation}
\rho_{G,+}(\omega) = \frac{P_+(i\omega\! +\! R_{+,E})\! +\! P_E(\Gamma_{1,1}\!+\!\Gamma_{1,2})}{(i\omega\!+\!R_{G,+})(i\omega\! +\! R_{+,E})\! -\! (\Gamma_{1,1}\! +\!\Gamma_{1,2})\Gamma_{\rm{inc}}},
\end{equation}
\begin{equation}
\rho_{-,E}(\omega) = -\frac{P_E(i\omega\! + \!R_{G,-})\! +\! P_- \Gamma_{\rm{inc}}}{(i\omega\!+\!R_{-,E})(i\omega \!+\! R_{G,-})\!-\! (\Gamma_{1,1}\! -\!\Gamma_{1,2})\Gamma_{\rm{inc}}},
\end{equation}
and $\rho_{+,E}(\omega) = \big[P_E + \Gamma_{\rm{inc}}\rho_{G,+}(\omega)\big]/\big[i\omega+R_{+,E}\big]$, $\rho_{G,-}(\omega) = \big[P_- - (\Gamma_{1,1}-\Gamma_{1,2})\rho_{-,E}(\omega)\big]/\big[i\omega+R_{G,-}\big]$. Here, $R_{+,E} = (3\Gamma_{1,1}+\Gamma_{1,2}+\Gamma_{\rm{inc}}+\Gamma')/2 + i\delta_{1,2}$, $R_{G,+} = (\Gamma_{1,1}+\Gamma_{1,2}+3\Gamma_{\rm{inc}} + \Gamma')/2 -i\delta_{1,2}$, $R_{G,-} = (\Gamma_{1,1}-\Gamma_{1,2}+3\Gamma_{\rm{inc}} +\Gamma')/2 + i\delta_{1,2}$, and $R_{-,E}=(3\Gamma_{1,1}-\Gamma_{1,2}+\Gamma_{\rm{inc}}+\Gamma')/2 -i\delta_{1,2}$. The values $P_E$, $P_+$, and $P_-$ refer to the steady state populations of the system eigenstates and are given by:
\begin{equation}
P_{\pm} = \frac{\Gamma_{1,1} +\Gamma_{\rm{inc}}+\Gamma' \mp\! \Gamma_{1,2}(1-\Gamma_{\rm{inc}}/\Gamma_{1,1})}{\Delta_D},
\end{equation}
and
\begin{equation}
P_E = \frac{\Gamma_{\rm{inc}}(1+\Gamma_{\rm{inc}}/\Gamma_{1,1}+\Gamma'/\Gamma_{1,1})}{\Delta_D},
\label{Ppm}
\end{equation}
with 
\begin{align}
    \Delta_D &=  3(\Gamma_{1,1}+\Gamma_{1,2}) + \frac{(\Gamma_{\rm{inc}}^2+\Gamma_{1,2}^2)}{\Gamma_{1,1}}\ \nonumber \\ &+ \frac{(\Gamma_{1,1}^2 -\Gamma_{1,2}^2)}{\Gamma_{\rm{inc}}}+\frac{\Gamma'(\Gamma_{1,1}+\Gamma_{\rm{inc}})^2}{\Gamma_{1,1}\Gamma_{\rm{inc}}}.
    \end{align}

In the regime of $\delta_{1,2} \gg \{\Gamma_{1,1},\Gamma',\Gamma_{1,2},\Gamma_{\rm{inc}}\}$, the spectrum can be further simplified to a simple form, as a sum of Lorentzians. Then, we  have:
\begin{equation}
\rho_{i,j}(\omega) = \frac{P_{j}}{i\omega+R_{i,j}},
\end{equation}
for $(i,j) = \{(G,+),(G,-),(+,E),(-,E)\}$. Note that in the weak excitation approximation, the terms $\rho_{+,E}(\omega)$ and $\rho_{-,E}(\omega)$ are zero. For a detector placed far away from the two coupled QDs, $g_{n,n'}(\omega)$ vary little from each other, and the total far-field spectrum takes on the simple form:
\begin{equation}
S(\omega) \approx \rm{Re}\big\{\rho_{+,E}(\omega) + \rho_{G,+}(\omega)\big\},
\end{equation}
which consists of a peak at $\omega = -i\delta_{1,2}$ with FWHM $3\Gamma_{1,1}+\Gamma_{1,2}+\Gamma_{\rm{inc}}+\Gamma'$, and a peak at $\omega = +i\delta_{1,2}$ with FWHM $\Gamma_{1,1}+\Gamma_{1,2}+3\Gamma_{\rm{inc}} + \Gamma'$, where only the latter peak remains in the linear spectrum. Thus for $\Gamma_{1,1} > \Gamma_{\rm{inc}}$, the FWHM of the peak at $-i\delta_{1,2}$ is broader than the one at $i\delta_{1,2}$ by $2(\Gamma_{1,1} - \Gamma_{\rm{inc}})$, and vice-versa for $\Gamma_{1,1} < \Gamma_{\rm{inc}}$. Similarly, the ratio of the spectral weights (integrated intensity) of the peak at $-i\delta_{1,2}$ to the one at $i\delta_{1,2}$ can be found as $P_E/P_+$, which is (approximating $\Gamma_{1,1} \approx \Gamma_{1,2}$ for simplicity):
\begin{equation}
\frac{P_E}{P_+} \approx \frac{1}{2}\left (\frac{1+(\Gamma_{\rm{inc}}+\Gamma')/\Gamma_{1,1}}{1+\Gamma'/(2\Gamma_{\rm{inc}})}\right ),
\end{equation}
which transitions at the $\Gamma_{1,1} = \Gamma_{\rm{inc}}$ point.
Note that for $\Gamma_{1,1} - \Gamma_{1,2} < \Gamma_{\rm{inc}}$, the weak excitation approximation does not accurately predict the emission spectrum, as the $\ket{\Psi_-}$ eigenstate becomes optically dark, and thus $P_E$ will be excited substantially even for a weak pump. 

To compare with a weak excitation approximation, we can derive the previous equations assuming the system to be only the set of $\ket{G}$, $\ket{\Psi_+}$, $\ket{\Psi_-}$ eigenstates. Then $\rho_{\pm,E}=0$, and we have the weak excitation results for $\rho^w_{G,\pm}(\omega)$:
\begin{equation}
    \rho^w_{G,\pm}=\frac{P^{\rm w}_{\pm}}{i(\omega \mp \delta_{1,2}) +\frac{1}{2}(\Gamma_{1,1} \pm \Gamma_{1,2} + 2\Gamma_{\rm{inc}}+\Gamma')},
    \label{eq:wea}
\end{equation}
where 
\begin{equation}
P^{\rm w}_{\pm} = \frac{\Gamma_{1,1}+\Gamma'\mp\Gamma_{1,2}}{2(\Gamma_{1,1}+\Gamma')+(\Gamma_{1,1}^2+\Gamma_{1,1}\Gamma'-\Gamma_{1,2}^2)/\Gamma_{\rm{inc}}}.
\end{equation}
The effects of the nonlinearities are made clear by comparing to Eq.~\ref{Ppm}.

\section{Results}
\label{sec:results}

\subsection{Numerical calculations of the nanowire photon Green function, field profiles, Purcell factors, and beta factors}

\begin{figure}[htp]
    \centering
    \includegraphics[width = 0.65\columnwidth]{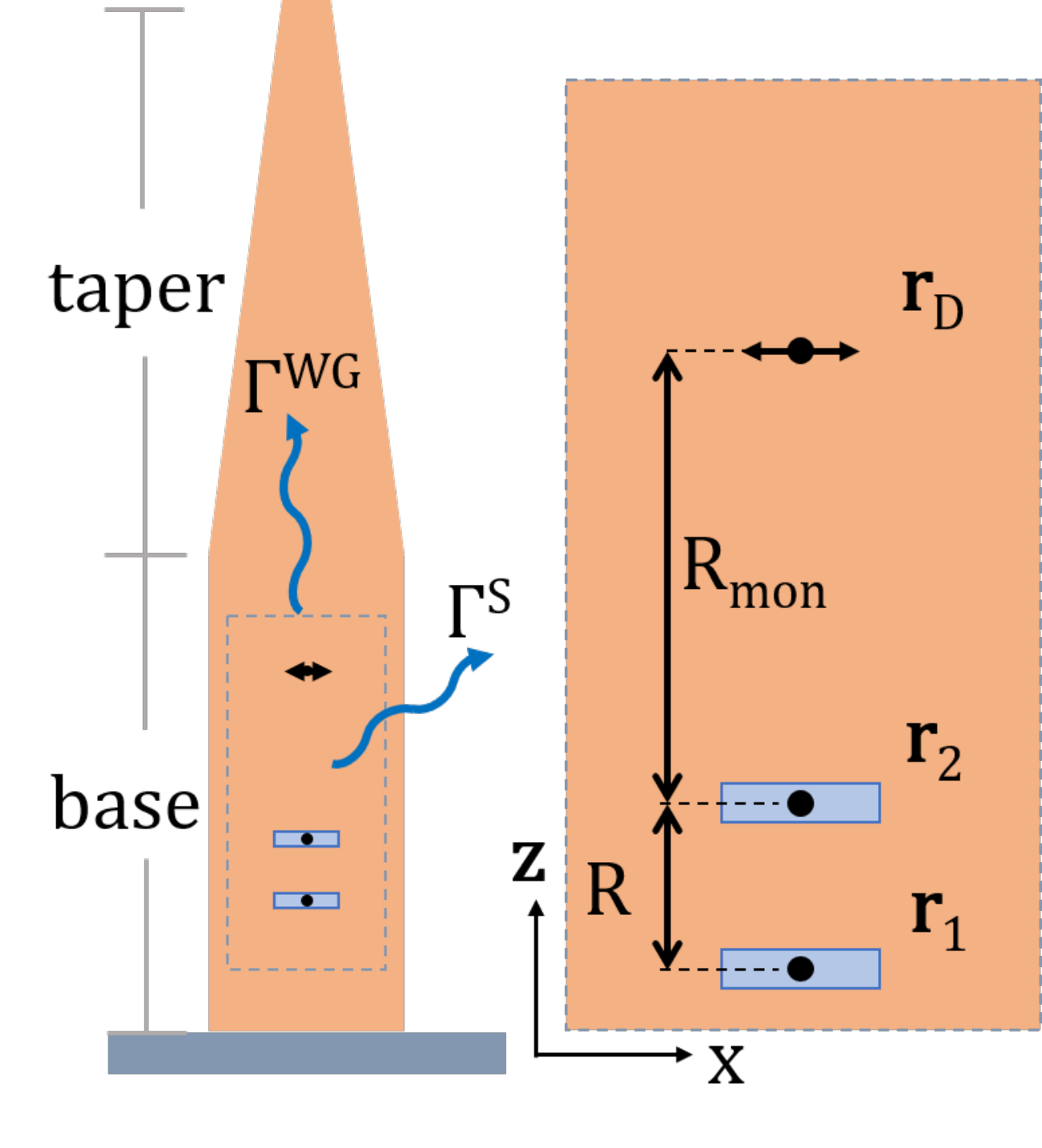}
    \caption{Schematic side-view of a nanowire where two QDs with $x$/$y$ dipole moments are separated only in $\mathbf{\hat{z}}$ and a detection point is located at $\mathbf{r}_D$; {$R_{\rm mon}$ is defined as the distance between the detector position and the second/upper QD}. For small separations, the GF is dominated by the homogeneous solution. To help understand the underlying physics, we will connect to point detectors as a function of height, but in practice these would be outside of the waveguide geometry, e.g., captured by a fiber. The radiative decay rates into the WG, $\Gamma^{\rm WG}$, and out of the wire (side), $\Gamma^{\rm S}$, are labelled as well. }
    \label{fig:SpectraSetup}
\end{figure}

In Fig.~\ref{fig:SpectraSetup}, we first show a simple schematic of two
dots (point dipoles) in a nanowire, with a taper that is usually (experimentally) included
to maximize the coupling vertically to a fiber. Below, we will
consider calculations for the infinite wire as well as the
tapered nanowire, and point out any subtle differences. The electromagnetic response of a complex photonic system is usually not known analytically, yet semi-analytical models can often be adopted when guided by the full numerical solutions. Numerical approaches can return both the analytical modes as well as the numerically exact GF, within numerical precision. To do this, we use  FDTD methods \cite{Taflove2005} to perform a full 3D GF analysis \cite{Yao2010} using Lumerical FDTD \cite{LumericalSolutions}, a commercially available Maxwell equation solver.  This FDTD method uses a Cartesian grid (Yee cell \cite{Yee1966}) to discretize the system in space, solving the electromagnetic fields at each cell location in discrete steps forward in time.

\begin{table}[htp]
    \centering
    \begin{tabular}{|c c|}
    \hline
         $n_B$ &  3.37             \\
         $n_{\rm{eff}}$ (FDTD, best fit)  & 1.86, 1.905  \\
         $n_{\rm{g}}$ (FDTD, best fit) & 5.35, 5.25    \\
         $L_{\rm{eff}}$ (FDTD, best fit) & 155, 180~nm  \\
         $f_{0}$  & 324.1~THz \\
          $\hbar \omega_{0}$  & 1.34~eV  \\
         $r$  & 110~nm  \\
         $\beta$ & 0.9 \\
    \hline
    \hline
         height (base, taper) & 1~$\mu$m, 5.7~$\mu$m \\
         $r$ (base, taper) & 110~nm, 10~nm \\
         taper & 1$^{\rm{o}}$ \\
         dipole $\mathbf{r}_0$ & 260~nm $\hat{\mathbf{z}}$\\
         $n_{\rm{substrate}}$ & 3.37 \\
    \hline
    \end{tabular}
    \caption{Waveguide parameters used in the analytic GF, obtained from the mode solution of the wire waveguide in FDTD. The effective mode length is simply $L_{\rm eff}=A_{\rm eff}^{1/2}$ as defined by $A_{\rm eff} = {1}/{\epsilon_{\rm B}|{\bf e}_k({\bm \rho}_0)|^2}$, with $\int_{A_{\rm eff}}  \epsilon({\bm \rho})  |{\bf e}_k({\bm \rho})|^2d{\bm \rho} =1$, as discussed for Eq.~(\ref{eq:Gwg}).}  
    \label{Tab:WireProperties}
\end{table}

\begin{figure}[htp]
    \centering
    \includegraphics[trim = 0.5cm 0.75cm 1.5cm 1.5cm, clip=true,width=\columnwidth]{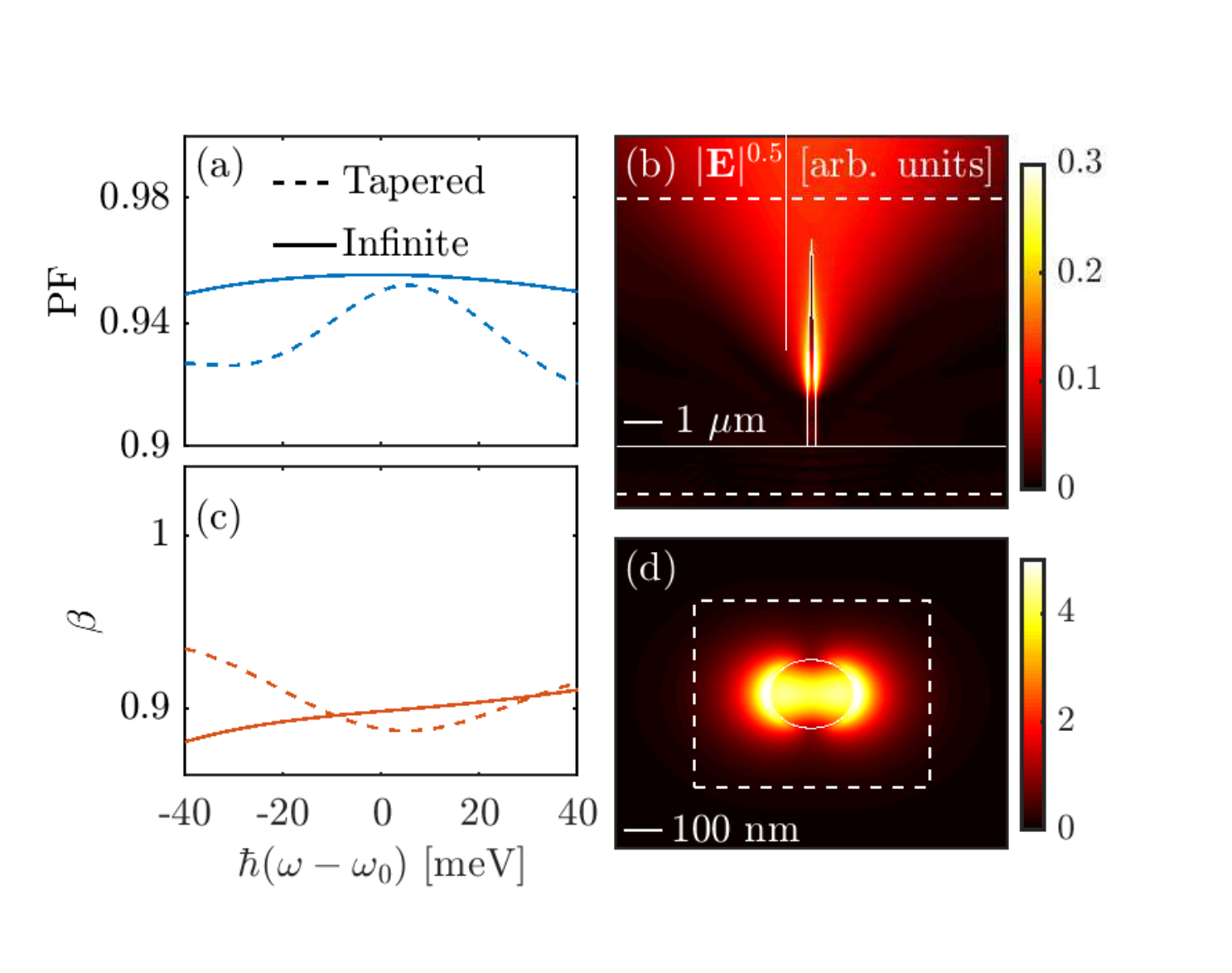}
    \caption{Purcell factor (a) and $\beta$ factor (c) for the tapered (dashed) and infinite (solid) nanowire with radius of 110~nm. The magnitude of the electric field in the (b) $x-z$ plane of the tapered wire and (d) in the $x-y$ plane of the infinite wire. The dashed white lines show the $z$-monitor locations/sizes used for the calculated $\beta$ factor given by Eq.~(\ref{eq:beta}). See text for more details.}
    \label{fig:BetaPF}
\end{figure}

To obtain the numerical GF in Eq.~(\ref{eq:GFanalytic}) using FDTD, we  solve the time-domain electric field response in the presence of a single dipole polarization source and open boundary conditions (perfectly matched layers, or PML) such that 
\begin{equation}
{G}_{{i,j}}(\mathbf{r}_1,\mathbf{r}_2;\omega) = \frac{{\rm FFT}[E_i(\mathbf{r}_1,t)]}{{\rm FFT}[P_j(\mathbf{r}_2,t)]},
\label{GFfdtd}
\end{equation}
where ${\rm E}_i$ is the $i^{\rm{th}}$ component of the electric field response at $\mathbf{r}_1$, 
${P_j}$ is the $j^{\rm{th}}$ component of the polarization response of the dipole at $\mathbf{r}_2$, and FFT is the fast Fourier transform from the time to the frequency domain. Care must be taken when calculating the GF using the Yee cell configuration, as the $x$, $y$, and $z$ components of the electric and magnetic fields are obtained at different locations within the cell; thus, the dipole and time monitors must be placed at the correct locations within the Yee cell corresponding to the polarization or field component of interest. As well, the real part of the total GF analytically  diverges as $\mathbf{r}_1 \rightarrow \mathbf{r}_2$, though FDTD gives a finite answer which is a volume averaged result over the mesh cell~\cite{VanVlack2012c}. For small separations, the size of the mesh cell is therefore of paramount importance, if an accurate real part of the GF is needed. As a result of the off-set locations for the field components in a Yee cell~\cite{Taflove2005} and mesh size dependence of the GF, there are also cross coupling terms (i.e. $xy$ for separations strictly in $z$) that appear in the FDTD method that do not appear in the analytical result of two point dipoles. These unintentional off-diagonal terms in the GF contribute to the total GF (and thus, the spectral splitting between two QDs), which may be important to consider for real QDs which are separated on the same order of magnitude as their lateral radii.   \\

One useful quantity to examine in such structures is the $\beta$ factor, which can be described as a measure of how much of the light is coupled into a particular optical mode of interest (i.e. of the NW or of a output coupling fibre above the NW). As shown in Fig.~\ref{fig:BetaPF}, this can be calculated by measuring the transmission through a power monitor normal to the axial direction of the NW ($z$, here):
\begin{equation}
    \beta = \frac{T_{{z}}^{\rm{+}} + T_{{z}}^{\rm{-}}}{\rm{PF}}, 
    \label{eq:beta}
\end{equation}
where $T_{{z}}^{\rm{+/-}}$ is the Poynting vector power transmission through the top/bottom $z$ plane with a specific size/location, and PF is the Purcell factor (i.e. local field enhancement due to the environment),
\begin{equation}
    {\rm PF}_{{i}}(\mathbf{r}_0, \omega) = \frac{{\rm Im}[{G}_{{\rm tot}}|_{{i,i}}(\mathbf{r}_0,\mathbf{r}_0;\omega)]}{{\rm Im}[{G}_{{\rm hom}}|_{{i,i}}(\mathbf{r}_0,\mathbf{r}_0;\omega)]},
    \label{eq:pf}
\end{equation}
where the homogeneous GF, ${G}_{\rm{hom}}|_{{i,i}}$, is given by Eq.~\ref{eq:GFanalytic}, and the total GF, ${G}_{\rm{tot}}|_{{i,i}}$, is the full response given by Eq.~\ref{GFfdtd} at the location of the dipole emitter.

\begin{figure*}[ht]
    \centering
    \includegraphics[width = 0.99\textwidth]{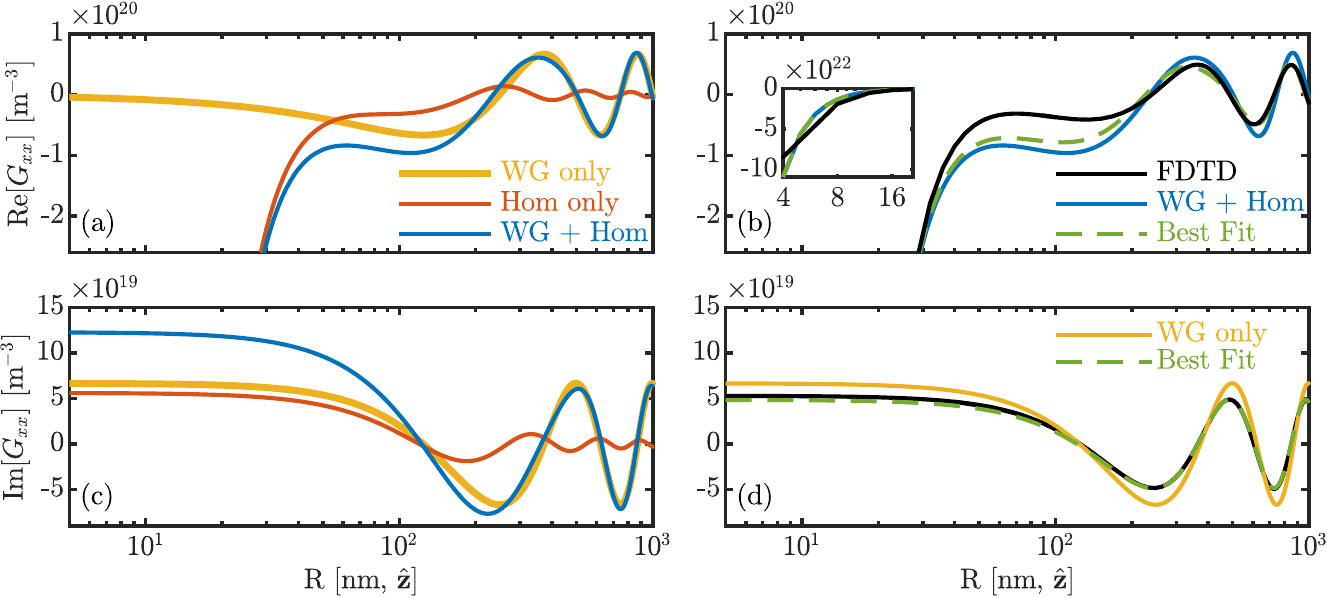}
    \caption{A direct comparison of the two-point GF as a function of separation ($xx$ component), $R$, along the axis of a nanowire using the waveguide (Eq.~\ref{eq:Gwg}), homogeneous (Eq.~\ref{eq:GFanalytic}), as well as the sum of the two GFs at $\hbar \omega_0$ = 1.34~eV ($\lambda_0$ = 925~nm). The waveguide parameters are given in Table~\ref{Tab:WireProperties}. Full FDTD simulations of {an infinite} NW are shown to examine the  approximation that ${\rm Re}[{\bf G}({\bf r}_1,{\bf r}_2)] = {\rm Re}[{\bf G}_{\rm{hom}}({\bf r}_1,{\bf r}_2)] + {\rm Re}[{\bf G}_{\rm{WG}}({\bf r}_1,{\bf r}_2)]$
    and ${\rm Im}[{\bf G}({\bf r}_1,{\bf r}_2)] = {\rm Im}[{\bf G}_{\rm{WG}}({\bf r}_1,{\bf r}_2)$], unless $\mathbf{r}_1 = \mathbf{r}_2$, in which case ${\rm Im}[{\bf G}({\bf r}_1,{\bf r}_1)] = {\rm Im}[{\bf G}_{\rm{WG}}({\bf r}_1,{\bf r}_1)] + (1 - \beta){\rm Im}[{\bf G}_{\rm{hom}}({\bf r}_1,{\bf r}_1)]$ (see text). } 
    \label{fig:compareGF}
\end{figure*}

Using the dipole excitation approach, the finite tapered NW and infinite NW which are examined numerically using FDTD have a radius of 110~nm and index of refraction of 3.37. The 2D mode solver in Lumerical FDTD was used to obtain the infinite wire's modal properties at the frequency of interest, which are summarized in Table~\ref{Tab:WireProperties}. The $\beta$ factor as well as the PF were obtained using a full 3D simulation of the infinite and finite tapered NW geometries (Eq.~\ref{eq:beta}), and are summarized in Fig.~\ref{fig:BetaPF}. It is important to highlight that the $\beta$ factor, in this definition, is highly sensitive to the size and position of the transmission monitor. For the infinite NW, the monitor was placed at $\pm$2\,$\rm{\mu m}~ \hat{\mathbf{z}}$ from the dipole emitter located in the center of the NW, and span 300~nm in $x$ and $y$ such that the mode is appropriately captured with minimal contributions from the leaky scattered light from the NW. However, for the finite tapered NW, the transmission monitors are placed at -1.500~$\rm{\mu m}$ and +8$\rm{\mu m}$ (wire spans from 0 to 6.7~$\rm{\mu m}$) and spans from -5 to +5~$\rm{\mu m}$ in $x$ and $y$. The reason for the large $x-y$ span is that the light would be experimentally collected via some type of fibre placed above/below the NW. Figure~\ref{fig:BetaPF} includes transmission data collected above the NW (top) as well as collected within the substrate (bottom) to get the total transmission, even though experimentally only the top would be collected. By symmetry, the infinite wire splits the transmission equally between the forward and backward $z$-directions, but due to symmetry breaking and the tapered design, the $\beta$ factor of the finite tapered wire is split such that 67.5\% of the light propagates upward toward a fibre and 32.5\% propagates downward into the substrate. Not only is the total $\beta$ factor improved in a tapered design, but the percent directed upward is enhanced. \\

The value of $\beta$, as defined above, is used with the PF to calculate the radiated emission rate into the WG and scattered out the side of the WG, $\Gamma^{\rm WG}$ and $\Gamma^{\rm S}$, respectively. The $\beta$-factor for the WG and the scattered light is then defined by, 

\begin{equation}
    \beta^{} = \frac{\Gamma^{\rm WG}}{\Gamma^{\rm WG}+\Gamma^{\rm S}},
\end{equation}
\begin{equation}    
    \beta^{\rm S} = \frac{\Gamma^{\rm S}}{\Gamma^{\rm WG}+\Gamma^{\rm S}}, 
\end{equation}
\noindent such that, 
\begin{equation}
    \Gamma^{\rm WG} = {\rm{PF}}\cdot\Gamma_{0}\beta,
\end{equation}
\begin{equation}    
    \Gamma^{\rm S} = {\rm{PF}}\cdot\Gamma_{0}(1 - \beta), 
\end{equation}

\noindent where ${\bf G}_{\rm hom}(\mathbf{r}_0,\mathbf{r}_0;\omega)$, $\beta$,
and the PF are determined via Eqs.~(\ref{eq:ghom}), (\ref{eq:beta}) and (\ref{eq:pf}), respectively, and $d=d_1$ (we also assume $d_1=d_2$ below). The values calculated here are consistent with those reported experimentally under similar conditions \cite{haffouz_bright_2018}.  

Using the FDTD mode solution as a guide, Fig.~\ref{fig:compareGF} directly compares the real and imaginary components of the analytic GF for three scenarios: ($i$) WG only, ($ii$) homogeneous only, and ($iii$) WG + homogeneous. The GF is calculated at the center frequency, $f_0$, for different separations along the axial direction of the wire, $R$, between 4 and 1000~nm; this spatial range adequately captures features from the near, intermediate, and far field contributions of the GF. The left hand panels, (a) and (c), show the three GF options calculated using Eqs.~(\ref{eq:Gwg}) and (\ref{eq:GFanalytic}) and the FDTD computed values summarized in Table~\ref{Tab:WireProperties}. It is evident that over the entire range of $R$, the real and imaginary GF of our nanowire is best described analytically by the WG+homoegenous and the WG only, respectively. The right hand panels, (b) and (d), compare the best analytic GF to the FDTD calculated GF for the infinite wire (black solid line), as well as the best analytic fit which is determined by varying $A_{\rm eff}$, $n_{\rm g}$, and $n_{\rm eff}$ within $\approx$15\% of the FDTD calculated values to minimize:

\begin{equation}
    \nonumber
\sum_{R = 4 {\rm nm}}^{1000 {\rm nm}} \bigg\vert\frac{{\rm Im}[G_{xx}^{\rm WG}(R)] - {\rm Im}[G_{xx}^{\rm FDTD}(R)]}{{\rm Im}[G_{xx}^{\rm FDTD}(R)]}\bigg\vert.
\end{equation}

The period of oscillation is affected only by $n_{\rm eff}$, thus, there is a clear optimization value for this variable. However, the amplitude of $G^{\rm WG}$ is proportional to $A_{\rm eff}/n_{\rm g}$, thus, there are infinitely many solutions to this optimization problem. We therefore used the pair of optimal values that best represented the original values of $n_{\rm g}$ and $A_{\rm eff}$. Note that we did not perform separate optimizations for the real and imaginary components of the GF due to the fact that both components should physically describe the same photonic system. \\

\begin{figure}[ht]
    \centering
    \includegraphics[trim = 0.5cm 0cm 2.35cm 0.15cm, clip=true, width=\columnwidth]{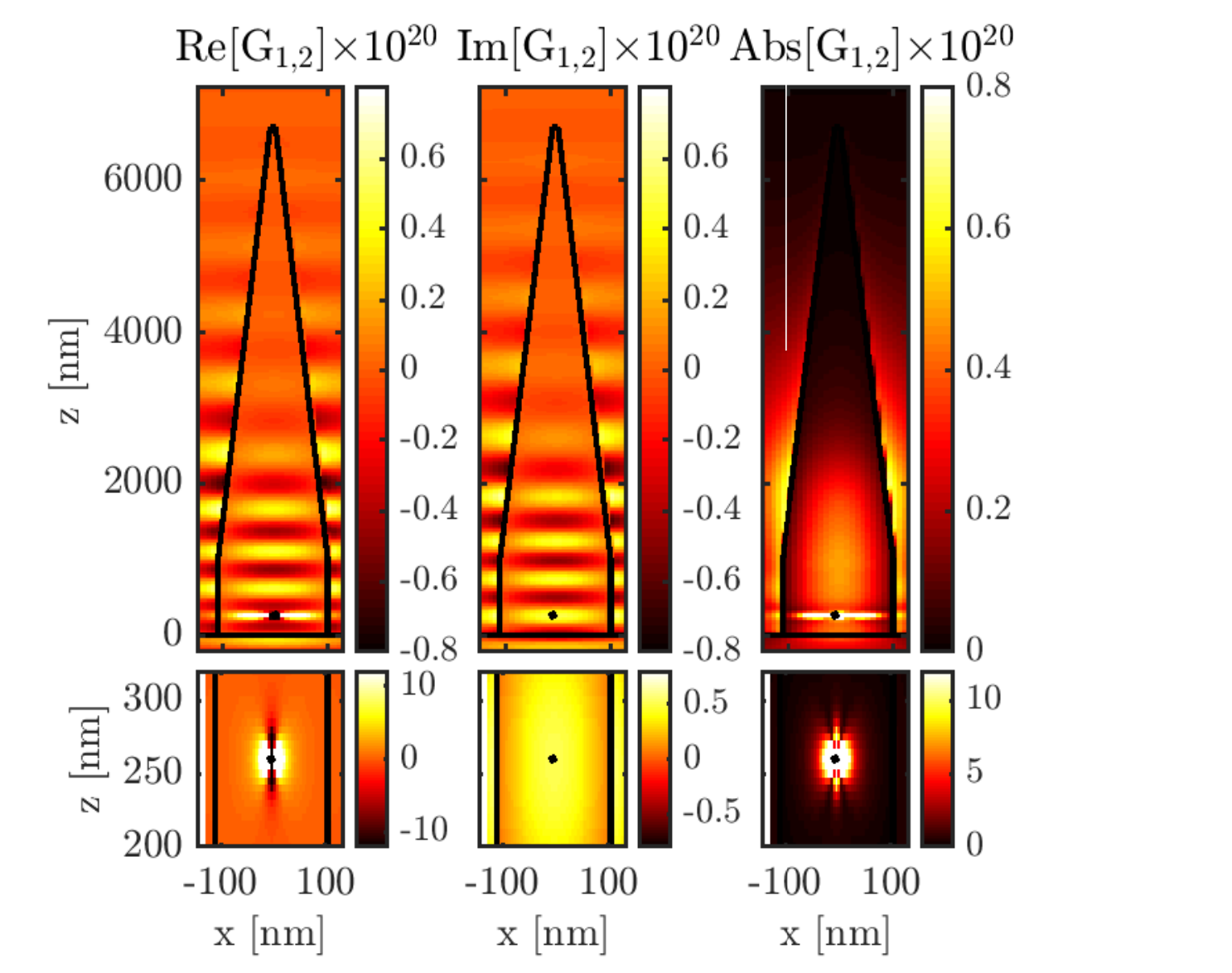}
    \caption{$G_{xx}$ of the nanowire taper, in units ${\rm m}^{-3}$; see text for details of the wire geometry and parameters.}
    \label{fig:GFcolormap}
\end{figure}

\begin{figure}[tbh]
	\centering
	\includegraphics[width=\columnwidth]{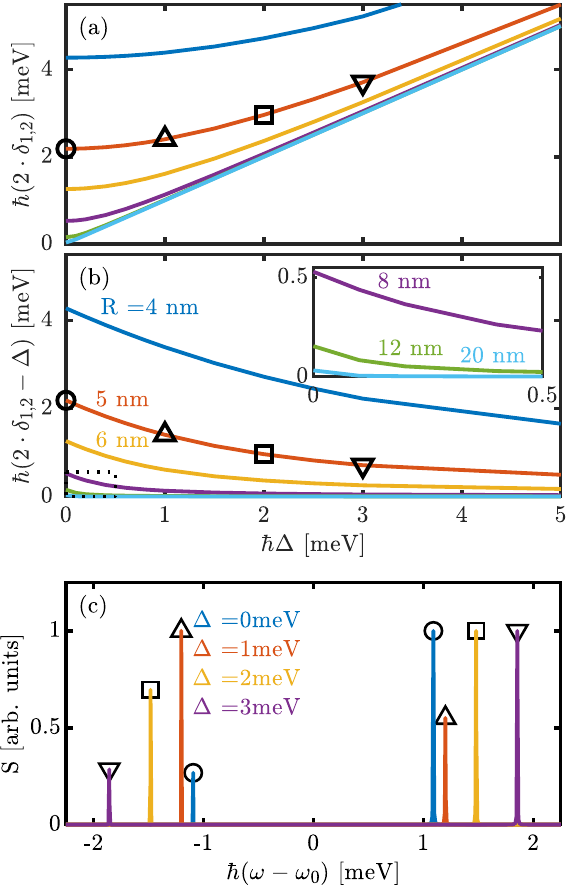}
	\caption{(a-b) Splitting as a function of detuning and dot separation for $d$ = 50~D, shown in two different ways. (c) Spectra for $R$=5~nm, $d$=50~D, and initial condition $\ket{\Psi_1}$ as a function of detuning. {The detection point, $\mathbf{r}_D$ is located 1000~nm~$\hat{\mathbf{z}}$ above the first/lower QD.} }
	\label{fig:detuningR}     
\end{figure}


The FDTD solution for the tapered NW design is not shown in Fig.~\ref{fig:compareGF}, since the infinite wire is the system that will be used in the rest of this work; however, we do present all of the components of $G_{\rm tot}|_{x,x}(\mathbf{r}, \mathbf{r}^{\prime},\omega_0)$ for the tapered NW in Fig.~\ref{fig:GFcolormap}. This data is represented as a 2D spatial contour map which illustrates qualitatively similar behaviour to the infinite wire, where the real part of GF rapidly increases as $\mathbf{r}\rightarrow\mathbf{r}^{\prime}$ and is then dominated by the WG contributions that create the oscillatory behaviour in the axial direction. The dipole in the tapered NW FDTD simulation is placed at an antinode, determined using a sweep PF calculations as a function of $R \hat{\mathbf{z}}$, to obtain the best results.\\

\subsection{Linear  spectrum from excited quantum dots in vacuum using the Green function approach}

In this section, we examine the emitted spectrum, $S(\omega)$, as defined by Eq.~\ref{eq:spectrum} for a variety of parameters including cross coupling in the GF, initial conditions, QD bare resonance detuning ($\Delta$), background broadening (through a complex $\omega_0$), and detection position. Note that most features of the spectral features can also be explained in terms of the molecular eigenstates as described by the   master equation model discussed in Sections \ref{sec:me} and \ref{sec:meresults}.     

Theoretically, we expect the QD resonance to split into two distinct resonances centered around $\hbar\omega_0$ and separated by {2}$\hbar\delta_{1,2}(\omega_0)$ (Eq.~\ref{eq:delta}), which is proportional to $1/R^3$ for small QD separations, as well as to the dipole moments of the QDs ($d^2$ if $d_1 = d_2$). For example, for $d_{1/2} = 50$~D (in $\hat{\mathbf{x}}$ {or} $\hat{\mathbf{y}}$) and $R$ = 5~nm, the expected splitting would be approximately 2.2~meV. These parameters will be considered our base example for all studies unless otherwise noted. Note that
the spectral splitting with double in value
if  both $\hat{\mathbf{x}}$ {and} $\hat{\mathbf{y}}$ dipole moments are considered.

First, let us consider detuning between the QDs, such that $\omega_{1/2} = \omega_0 \pm \Delta/2$. Figure~\ref{fig:detuningR} shows how the splitting changes as a function of detuning for a range of QD separations from 4~nm to 20~nm, as well as the spectra for $R$=5~nm for various values of detuning. We see that for small separations, the increase in splitting due to detuning is relatively small compared to the initial splitting and is non-linear. If we look at various separations, the effect of detuning changes rapidly such that as $R$ increases the resonances simply become $\omega_0 \pm \Delta/2$ (i.e., the dots become ``uncoupled'');  within a 
rotating-wave and Markov approximation, the resonances become $\omega_0 \pm \sqrt{\delta_{1,2}^2+{\Delta^2}/{4}}$. Importantly, the splitting for very close QDs ($R<$10~nm) is robust to the detuning that may be present even for nominally identical experimental dots. 

\begin{figure}[th]
    \centering
    \includegraphics[width = \columnwidth]{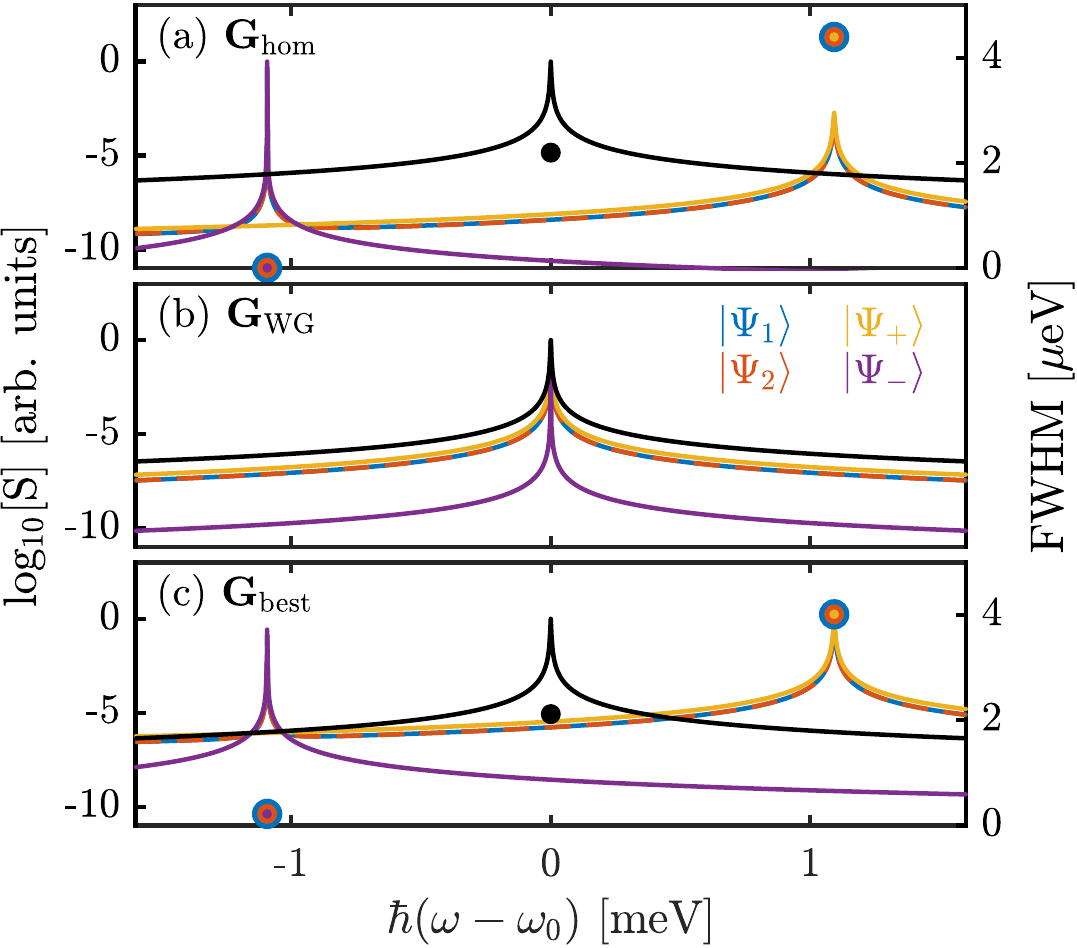}
    \caption{A direct comparison of the analytic spectrum, $S$, given by Eq.~\ref{eq:spectrum} at a detection point $\mathbf{r}_D$ = 1000~nm $\mathbf{\hat{z}}$ and dot separation of $\mathbf{R}$ = 5~nm $\mathbf{\hat{z}}$. The FWHM of each resonance (except for WG only, b) is shown on the right axis, determined using a Lorentzian fit, matching the expected FWHM from Eq.~\ref{eq:fwhm}. The initial conditions for $\ket{\Psi(t=0)}$ are labelled as described in Table~\ref{Tab:InitialConditions}. For comparison, the single QD (black) solution is shown. $\mathbf{G}_{\rm{hom/WG}}$ apply for both the real and imaginary components, whereas $\mathbf{G}_{\rm{best}}$ refers to the best fit described in Fig.~\ref{fig:compareGF}.}
    \label{fig:compareS}
\end{figure}
\begin{figure}[th]
	\centering
	\includegraphics[width=\columnwidth]{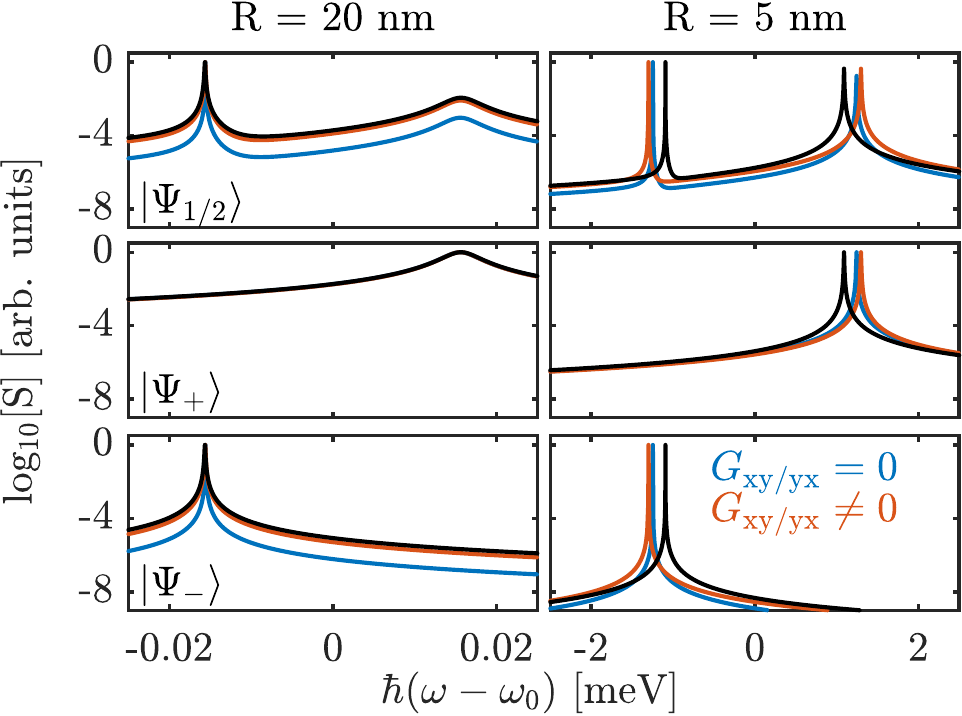}
	\caption{Analytic (black) and FDTD (colour) for two separations. Mesh size = 1~nm and d = 50~D. Note: we have used all four tensor components of the homogeneous GF. If the QDs have $x$-$xy$ {and} $y$-$yx$ coupling, then the splitting is precisely {double}. {The detection point, $\mathbf{r}_D$ is located 120~nm~$\hat{\mathbf{z}}$ above the first/lower QD.}}
	\label{fig:fdtd}
\end{figure}

Figure~\ref{fig:compareS} shows how the spectra changes for the four initial conditions defined in Table~\ref{Tab:InitialConditions}, as well as the full-width half-maximum (FWHM) of each of the resonances, using the three different GF models from Fig.~\ref{fig:compareGF}. The presence of dipole-dipole coupling, the conditions of $\ket{\Psi_{1/2}}$ (dashed lines) are analogous to either of the QDs being initially excited and the other in the ground state, where $\ket{\Psi_{+/-}}$ (solid lines) are linear combinations of $\ket{\Psi_{1/2}}$, and are the set of new eigenstates ($\ket{\Psi_{+/-}}$) of the system, similar to that of molecular dipole-dipole coupling and F\"{o}rster coupling \cite{tanas_entangling_2004,Unold2005, kim_exciton_2016, Stokes2018}. The analytic FWHM of the peaks will have values of
\begin{equation}
    \Gamma_{\pm} \approx \Gamma_{1,1} \pm \Gamma_{1,2} + \Gamma,
    \label{eq:fwhm}
\end{equation}
\noindent where $\Gamma$ is a possible additional single QD broadening.
{Note that $\Gamma_{1,2}$ accounts for incoherent coupling
of the dots mediated entirely through the waveguide mode, and specifically through the imaginary part of the waveguide
GF.}
Note that this is the same result given by the weak excitation regime in the master equation approach within the Markov approximation, given by Eq.~\ref{eq:wea}. One can easily verify this solution by fitting the spectra to a two-peak Lorentzian model. As expected, the real part of $G^{\rm WG}$ approaches zero for small separations, so no splitting is observed if only the WG is considered. The spectra are shown on the left axis and the FWHM on the right. Depending on the initial condition (i.e., initially excited molecular eigenstate), the spectra illustrates a super-radiant ($\omega>\omega_0$) and sub-radiant resonance ($\omega<\omega_0$), as shown by the FWHM relative to the single QD solution. When the WG is added to the homogeneous solution, the super- and sub-radiant solutions are still present, except with modified FWHM that reflect the additional broadening from the imaginary part of ${\bf G}^{\rm WG}$. 

Next, we will look at the possibility of cross-coupling in the GF. Figure~\ref{fig:fdtd} directly compares the analytical homogeneous GF with the FDTD homogeneous GF in the calculated spectrum with and without $x-y$ dipole cross coupling. As $R$ decreases, the GF is overestimated by FDTD (mesh of 1~nm used, here). This is a known effect due to finite meshing effects in the Yee cell \cite{van_vlack_finite-difference_2012}. As discussed in the previous section, it is very possible that at such close separations the dipole approximation breaks down due to the lateral dimensions of the QDs being  comparable to the vertical separation; in this case, cross coupling may contribute to the total GF, which is naturally captured in FDTD depending on the choice of mesh size. These effects are out of the scope of this paper, but are useful to keep in mind as a source of additional splitting. \\

\begin{figure}[!t]
	\centering
	\includegraphics[width=\columnwidth]{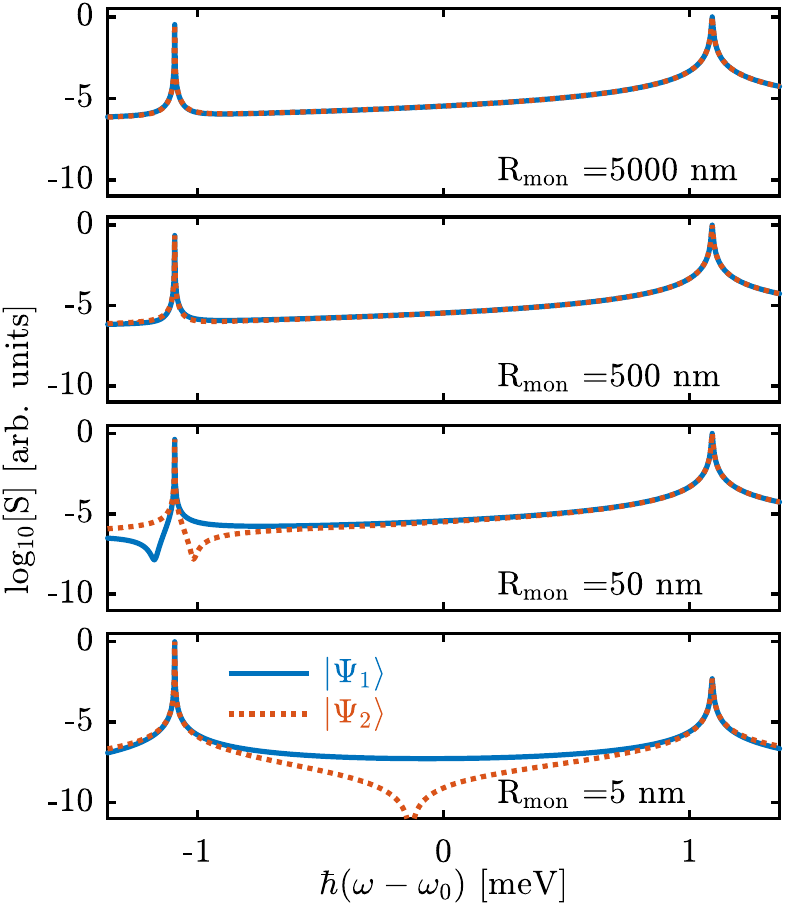}
	\caption{Analytic spectra as a function of detector position ($\rm R_{\rm{mon}}${, see Fig.~\ref{fig:SpectraSetup}})  where $r_D$ is directly above the QDs in the axial direction of the NW. $R$ = 5~nm and d = 50~D.\vspace{0.1cm}}
	\label{fig:detectordistance}
\end{figure}


As seen in Figs.~\ref{fig:compareS} and \ref{fig:fdtd}, $\ket{\Psi_1}$ and $\ket{\Psi_2}$ produce identical spectra. However, due to the asymmetry between the locations of $\mathbf{r}_1$, $\mathbf{r}_2$, and $\mathbf{r}_D$, this is not strictly true. In Fig.~\ref{fig:detectordistance}, we show that as the detection point moves closer to the QD pair in the axial direction, this asymmetry between the two states becomes evident, although only visible on a log$_{10}$ scale. For all other calculations, the detection point is taken to be 1000~nm such that the two states are effectively symmetric. This is understandable given that $\vert \mathbf{G}^{\rm{hom}}_{\rm 1,D}$-$\mathbf{G}^{\rm{hom}}_{\rm 2,D}\vert\rightarrow0$ as $\mathbf{r}_D\Rightarrow\infty$. \\


\subsection{Nonlinear spectra with incoherent pumping using a master equation solution}\label{sec:meresults}

\begin{figure}[htp]
    \centering
    \includegraphics[width=\columnwidth]{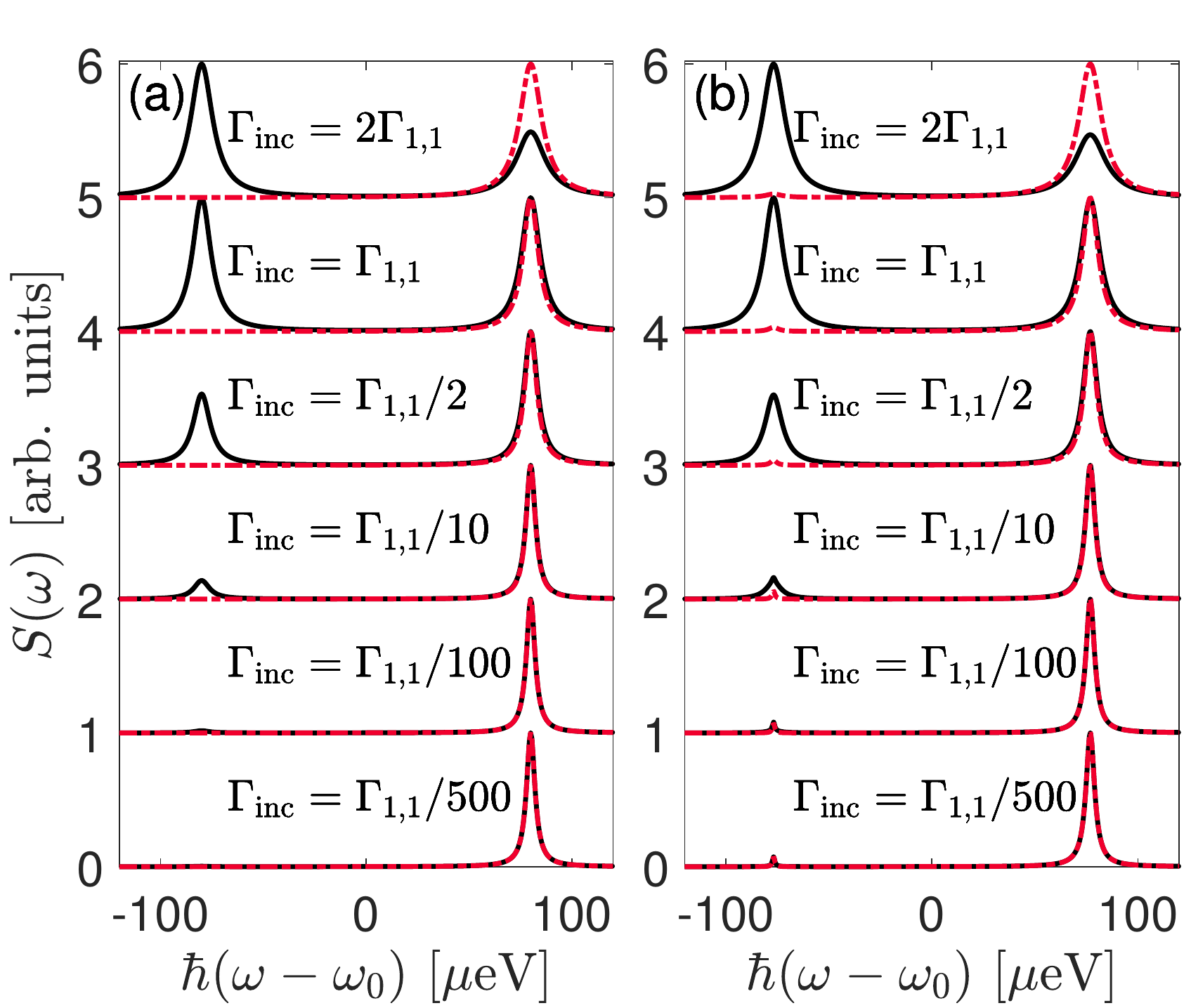}
    \vspace{-0.4cm}
    \caption{Emission spectra for incoherently pumped QDs for two sets of parameters using the full master equation solution (solid black) and the weak excitation approximation solution (dash-dotted red). For (a), $\hbar\Gamma_{1,1} = 2 \ \mu \rm{eV}$, $\Gamma_{1,2} = 0.999\,\Gamma_{1,1}$, $\hbar\delta_{1,2} = 80 \ \mu \rm{eV}$, and all the $g_{n,n'}(\omega)$ are assumed equal and independent of $\omega$, such that all transitions involving the $\ket{\Psi_-}$ eigenstate destructively interfere. In (b), we use the ``best fit'' parameters and GF, with $R= 12 \ \rm{nm}$, $r_D-r_1 = 1 \ \mu\rm{m}$, and $d = 50 \ \rm{D}$. For both (a) and (b), $\Gamma' = \frac{1}{2}\Gamma_{1,1}$.}
    \label{fig:NonLinear}
 \vspace{0.2cm}
    \includegraphics[width=\columnwidth]{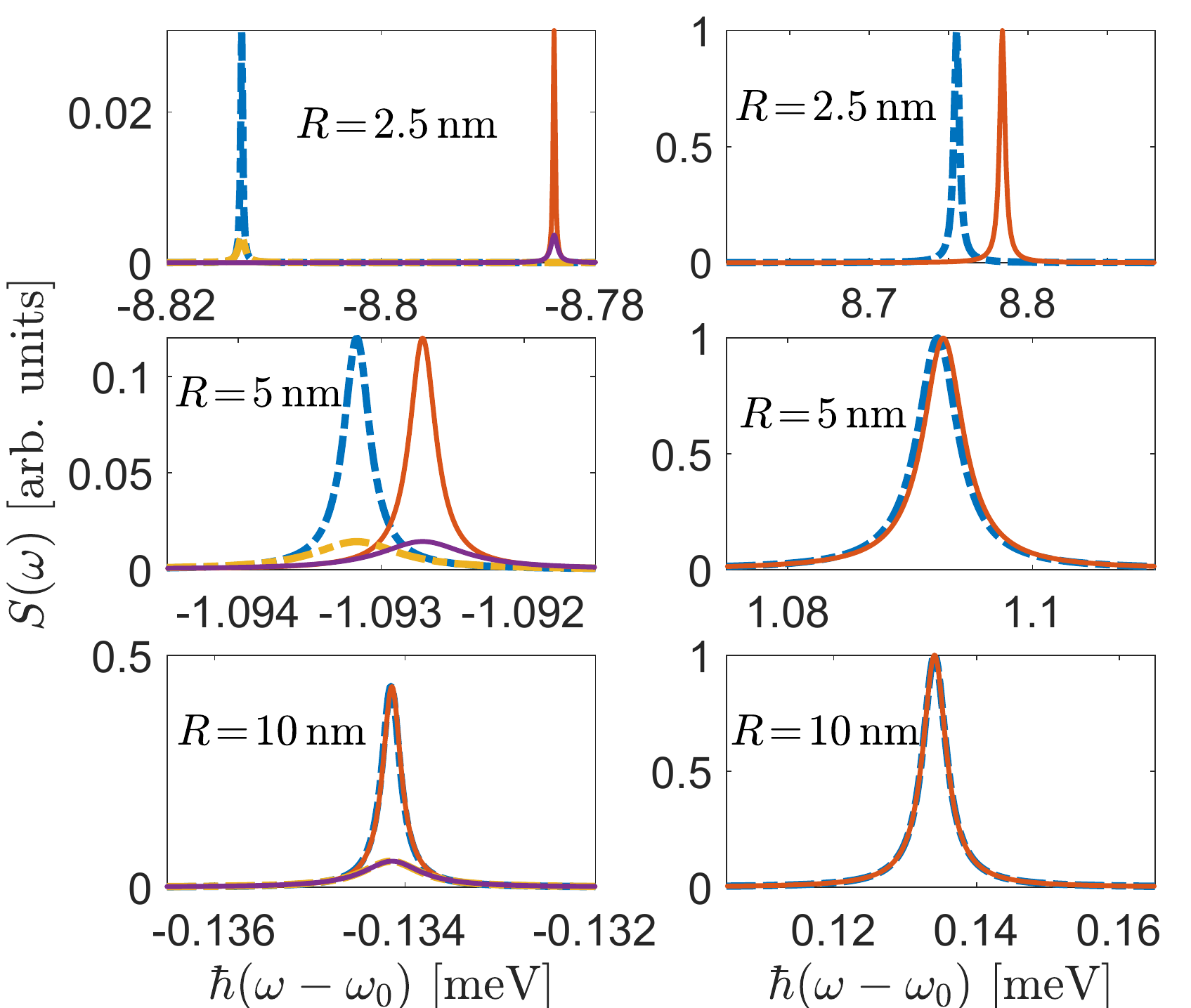}
        \vspace{-0.4cm}
    \caption{Comparison of linear emission spectrum calculated with the weakly-pumped ($\Gamma_{\rm{inc}} =10^{-5}\Gamma_{1,1}(\omega_0)$) master equation solution of Sect.~\ref{sec:me} (solid lines) with the GFn approach of Sect.~\ref{sec:theory} (dash-dotted lines). Here, $r_D-r_1 = 100 \ \rm{nm}$, $d = 50 \ \rm{D}$, and the dot separation is varied. For clarity, the lower and higher frequency peaks are shown separately on the left and right panels, respectively. To compare additional broadening mechanisms, we let the numerical value of the master equation approach pure dephasing to be equal to the additional broadening in the GF approach ($\Gamma' = \Gamma$); for the blue and red curves, $\Gamma' = 0$, and for the purple and yellow, $\hbar\Gamma' =0.5 \ \mu \text{eV}$ (omitted on right panels for visual clarity). }
    \label{fig:Comparison}
\end{figure}

In this subsection, we discuss the results from the master equation approach, which allows for the nonlinear quantum regime to be studied.
{Such effects are now beyond what could be captured by a classical Maxwell solution.}
As discussed in Sect.~\ref{sec:me}, the emission spectrum of two coupled QDs in the regime where $\delta_{1,2} \gg \{\Gamma_{1,1},\Gamma',\Gamma_{1,2},\Gamma_{\rm{inc}}\}$ consists of four superimposed Lorentzians with center frequencies $\omega_0 \pm \delta_{1,2}$. These peaks form from optical transitions between the dressed states of the system, where transitions involving $\ket{\Psi_-}$ are rendered mostly dark by interference in the far field. In the linear spectra (the regime of the weak excitation approximation), two of these peaks are non-zero, while in the nonlinear regime the $\ket{E}$ state population becomes substantial and the other transitions appear in the spectrum. For strong pumping, this can cause the spectral weights, as well as the linewidths of the most dominant peaks in the spectrum to flip in magnitude (see Sect.~\ref{sec:me}).  In Fig.~\ref{fig:NonLinear}, we plot the master equation solution with and without a weak excitation approximation for two sets of parameters. In strongly coupled QDs, the features of nonlinearity can appear even for relatively weak pumping, as the $\ket{\Psi_-}$ state is very nearly optically dark, allowing for population buildup in this state and thus easier pumping to the $\ket{E}$ state.

In Fig.~\ref{fig:Comparison}, we plot the emission spectrum in the linear excitation regime for both the master equation and GF approaches, to investigate the validity of the Markov and rotating-wave approximations made in the derivation of the master equation. Notably, the effect of additional broadening in the Green function approach is shown to be nearly equivalent to adding a pure dephasing with the master equation, in both the spectral weights and degree of broadening. Except at very strong coupling strengths (small dot spatial separation), for which other effects such as electronic tunneling and the breakdown of the dipole approximation are likely  significant, the differences in the two approaches are negligible. Either approach is thus appropriate for modelling the dipole-dipole coupling in nanowire waveguides under these parameter regimes, with the master equation solution allowing for insight into the nonlinear excitation regime. Furthermore, a more accurate master equation for strong-coupling between dots can be derived by first including the Coulomb dipole-dipole interaction in the system Hamiltonian before tracing over the photonic reservoir~\cite{Stokes2018}. 


\section{Photoluminescence Experiments on quantum dot molecules in Indium Phosphide Nanowires}
\label{sec:exp}

\begin{figure}[htb]
\centering
\includegraphics[width=1\columnwidth]{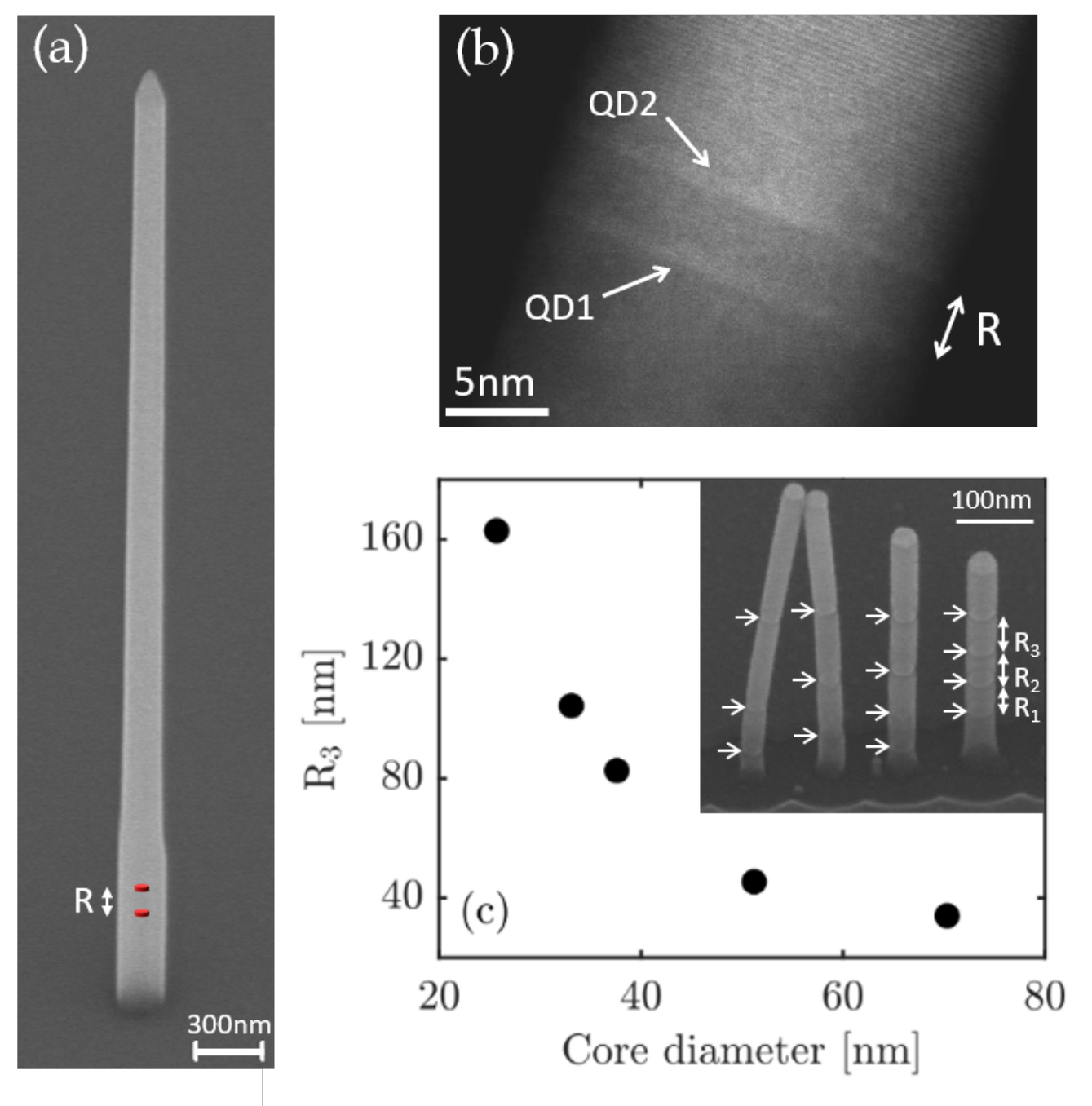}
\caption{(a) SEM image of a tapered InP nanowire waveguide. The position of the QD molecule in the waveguide is schematically indicated by the two red disks. (b) HRTEM image of the nanowire core showing two InAsP quantum dots separated by a 5 second InP spacer. (c) Dependence of $R$ on core diameter.  The inset shows a SEM image of 4 nanowires with different diameters. Each nanowire nominally contains 4 QDs, each separated by 60 seconds of InP growth. The QDs are delineated using a selective wet-etch and indicated by arrows. The smaller diameter nanowires are missing the first dot due to a diameter-dependent growth incubation time.}
\label{fig:SEM_TEM}
\end{figure}

To experimentally study waveguide-mediated coupling in QD molecues, we use bottom-up InP nanowires incorporating two InAsP QDs. The nanowires are grown using selective-area vapor-liquid-solid epitaxial growth on a patterned InP substrate\cite{Dalacu_NT2009}. Briefly, a Au catalyst is positioned in the center of a circular opening in a SiO$_2$ mask using a self-aligned lift-off process. Growth on such a substrate allows for independent control of the nanowire core (e.g. the quantum dots) defined by the Au catalyst and the waveguide defined by the oxide opening (see Refs. \onlinecite{Dalacu_NT2009, dalacu_ultraclean_2012} for details). A scanning electron microscopy (SEM) image of the nanowire waveguide and a high resolution transmission electron microscopy (HRTEM) of the nanowire core are shown in Fig.~\ref{fig:SEM_TEM}.

\begin{figure}[htb]
\centering
\includegraphics[width=0.99\columnwidth]{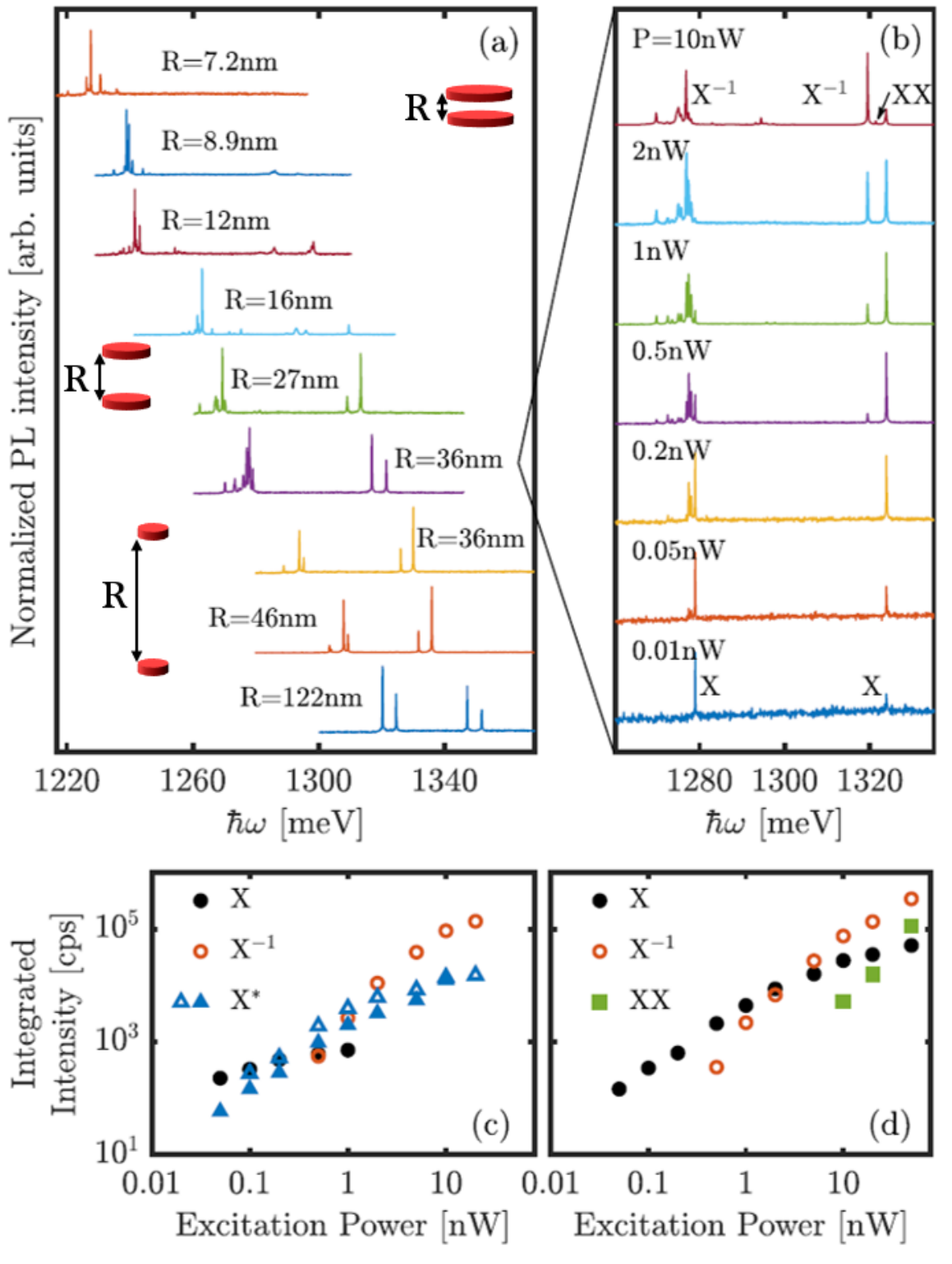}
\caption{(a) PL spectra of double QD nanowires with a 15 second spacer as a function of $R$ where $R$ is controlled using the core diameter-dependent growth rate. Excitation powers are 1\,nW except for $R$=7.2 and 122~nm, which are 5\,nW. (b) Power-dependent spectra from a nanowire with $R=36$\,nm, where the saturation power is 200~nW. Integrated PL intensities of the emission peaks from the high (c) and low (d) energy dot as a function of excitation power.}
\label{fig:NRCspectra2}
\end{figure}

\begin{figure}[th]
\centering
\includegraphics[trim = 1cm 3.5cm 4cm 0.5cm, clip=true, width=1\columnwidth]{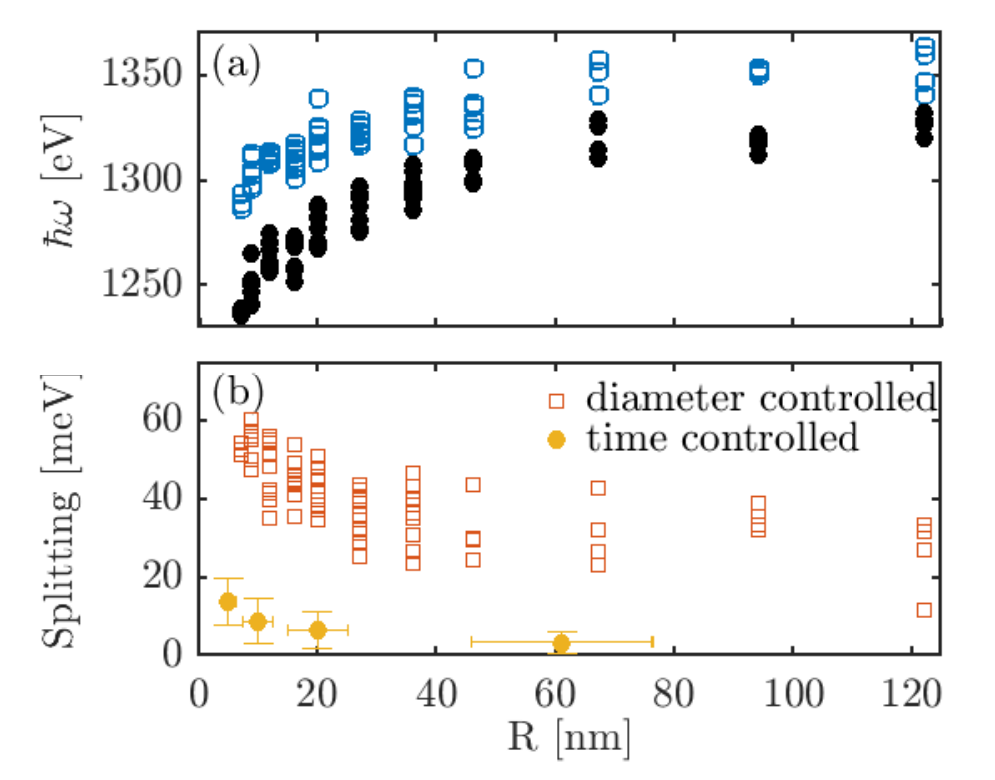}
\caption{(a) Emission energy of the $X^-$ transition in the high and low energy QD (diameter dependent growth only) and (b) their energy separation as a function of $R$. Excitation power is between 1-5~nW, as described in Fig.~\ref{fig:NRCspectra2}. The energies for the diameter (time) dependent data are extracted from spectra taken of up to 10 (75) nominally identical nanowires for each value of $R$. {The spectra for each sample of the time dependent data is shown in Fig.~\ref{fig:FullExp} in Appendix~\ref{appendix:data}, which includes the statistical analysis used to determine the error in splitting. The error in $R$ is determined experimentally.} }
\label{fig:NRCspectrasummary}
\end{figure}

\begin{figure}[th]
\centering
\includegraphics[trim = 0.3cm 0.75cm 1.2cm 3.5cm, clip=true,width = 1\columnwidth]{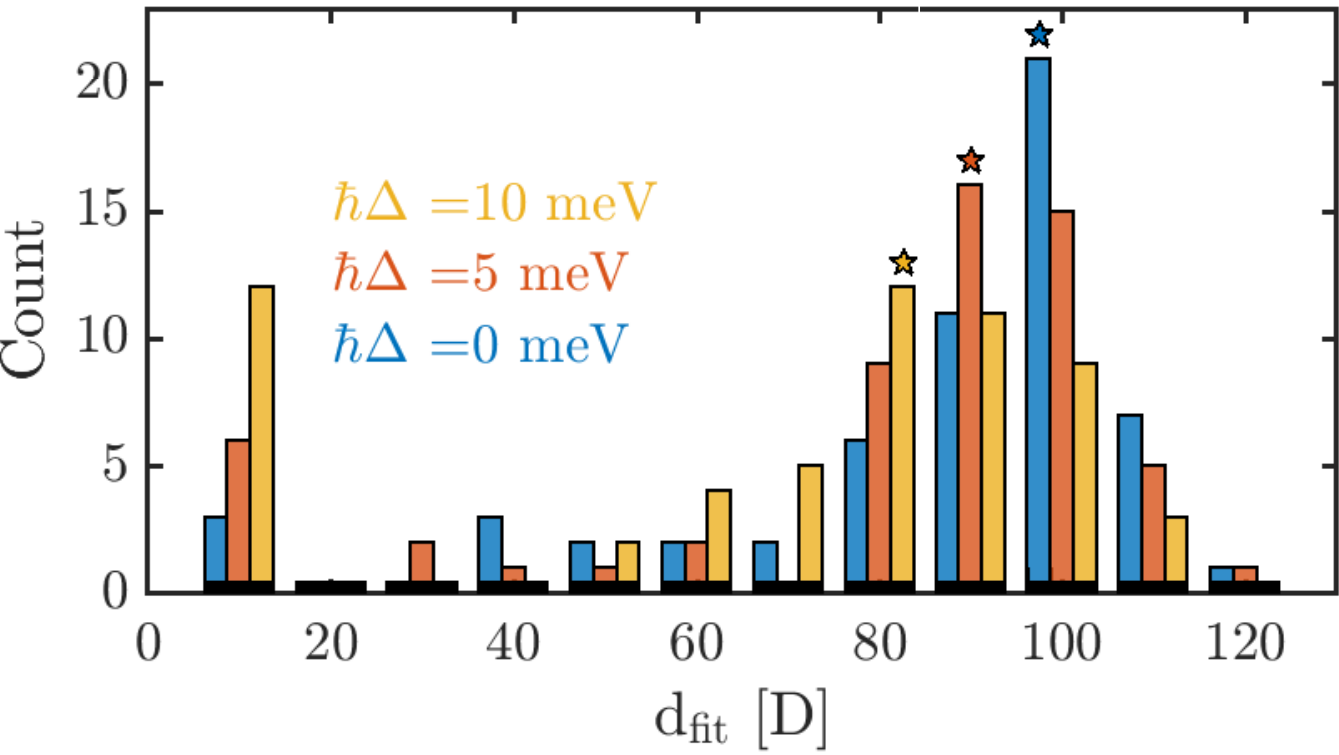}
\caption{Histogram of the dipole moment, $d_{\rm fit}$, used to fit to experimental spectral splitting for two QDs separated by $R$=5~nm, assuming a dipole moment oriented in both $\hat{\mathbf{x}}$ and $\hat{\mathbf{y}}$ (for results with only one dipole moment, multiply $d_{\rm fit}$ by $\sqrt{2}$); we neglect any cross-polarization mixing which would enhance the splitting and reduce the extracted
$d_{\rm fit}$. A total of 58 samples are used, and the statistics and individual spectra for each sample are given in Fig.~\ref{fig:FullExp} in Appendix~\ref{appendix:data}. Three nominal detunings (i.e., without dipole-dipole coupling) of 0, 5, and 10~nm (blue, red, and yellow, respectively) are shown. The star markers indicate the peak of each histogram. }
\label{fig:Exp-v-Theory}
\end{figure}

Quantitative analysis of the coupling is non-trivial due to the difficulty in experimentally determining $R$ in clad nanowires (e.g. in waveguides). As shown in Fig.~\ref{fig:SEM_TEM}(c) the nanowire growth rate is non-linear and diameter-dependent and there exists a growth incubation time which is also diameter-dependent. Unlike nanowire cores, the quantum dots in clad nanowires cannot be imaged with TEM to determine $R$. We rely instead on calibrated catalyst diameters and calculations based on a nonlinear growth model which is successful in predicting the heights of nanowire cores\cite{Dalacu_NT2009}. The accuracy of this method is limited by the process-related variation in the catalyst size ($\pm2$\,nm in a single run and slightly larger from run-to-run). 

We have investigated two approaches to controlling $R$. One relies simply on growing samples with nominally identical nanowires but with the growth time between incorporation of the first and second dot varied. In the second approach, we take advantage of diameter-dependence of the nanowire core growth rate. We grow double dot nanowires with controlled core diameters ranging from 16 to 70\,nm. Since the growth rate decreases with increasing diameter, devices with varying $R$ are obtained in a single growth as shown in Fig.~\ref{fig:SEM_TEM}(c). We note that, unlike the InP nanowire growth rate, the QD growth rate does not depend strongly on the indium flux provided to the growth system. The indium used for QD growth comes primarily from the excess indium in the gold particle, as described in \onlinecite{Froberg_NL2008}. This means that there is only a small dependence on the QD growth rate as a function on nanowire diameter. However, we still expect a diameter-dependence of the ground-state emission energy due to changes in the lateral confinement~\cite{Dalacu_APL2011}.

The $\mu$PL measurements were taken at 4K in a He-flow cryostat using above-band excitation at 633\,nm focused on the nanowires using a 50X (NA = 0.5) objective. The emission was collected through the same objective, dispersed using a grating spectrometer and detected with a nitrogen-cooled CCD. Experimental spectra for selected QD separations using the diameter-controlled approach are shown in Fig.~\ref{fig:NRCspectra2}\cite{khoshnegar_solid_2017}.
We observe a clear red-shift in the emission energies of the dots with decreasing $R$ (e.g. increasing core diameter) due predominantly to a decrease in lateral confinement as mentioned above. We also observe a gradual increase in the complexity of the low energy QD spectra and a decrease in the intensity of the high energy QD as $R$ is reduced.

Low-excitation PL spectra of QDs are typically multi-peaked. Recombination from different charge complexes may occur depending on background doping and Fermi level pinning. In our samples we typically observe emission from both neutral and singly-charged excitons. We assign the peaks based on excitation power-dependent measurements shown in Fig.~\ref{fig:NRCspectra2}. The spectra reduce to single a peak in the limit of low excitation power which we identify with the neutral exciton, $X$. For higher powers, i.e., $P=P_{\rm sat}/20$ where $P_{\rm sat}=200$\,nW is the excitation power required to saturate the transition in single dot nanowires, the brightest peak is typically the charged exciton, $X^-$. In Fig.~\ref{fig:NRCspectrasummary}, we plot the energies of the high and low $X^-$ peaks as well as the splitting as a function of $R$.


The energetic splitting of the $X^-$ peaks as a function of $R$ is consistent with the predicted behaviour described in Section C (see Figure~\ref{fig:detuningR}). However, it is clear from the non-vanishing splitting at large values of $R$ that the two emitters in isolation are not degenerate.  Such a non-degeneracy for dots grown under identical growth conditions indicates that the growth of the first dot affects the growth of the second dot. For example, this can be expected for closely spaced dots due the an arsenic tail from the first dot that increases the effective arsenic composition in the second dot, hence shifting the second dot to lower energy. For the data in Figure~\ref{fig:NRCspectrasummary} we have traced the non-degeneracy to a difference in the arsenic injection valve response time between the initial valve opening and subsequent openings. 

The additional measurements, where $R$ is controlled through growth time between the first and second dot, were made with the valve response time corrected. Also, PL spectra for each sample was collected from $>50$ nanowires to obtain a better measure of the splitting given the amount of scatter evident in Figure~\ref{fig:NRCspectrasummary}.

{All of the spectral data for the time-controlled samples are summarized in Figure~\ref{fig:FullExp} in Appendix~\ref{appendix:data}, where a total of four experimental QD separations are measured. The growth times between QDs are 15 seconds, 30 seconds, 1 minute and 3 minutes, and from our growth model, these times correspond to $R$-values of 5, 10, 20 and 61\,nm, respectively. The NWs are grown such that the diameter between each sample may vary, but the two QDs in each sample are nominally identical to each other (but not to the QDs in the next sample). Thus, we expect that there will be variance in the center frequency of the resonances, as mentioned above. For the purpose of this study, we will examine the 15~second (5~nm) data set more closely.} 

The extracted splittings are included in Fig.~\ref{fig:NRCspectrasummary}(b) where we observe an $R$-dependence similar to the previous data, but with an off-set of approximately -30~meV. The similarity in the observed spitting is consistent with the robust nature of the interaction in the presence of detuning (see Figure~\ref{fig:detuningR}). Although we cannot rule out an $R$-dependent detuning discussed above, supporting measurements made on closely spaced dots, including the observation of correlated emission between the two peaks\cite{khoshnegar_solid_2017} and a negative diamagnetic shift in magnetophotoluminescence\cite{Gaudreau_inprep}, strongly suggest a coupled dot system.  In the absence of a growth technique that guarantees zero detuning independent of $R$ (or a growth technique that guarantees a variance in the bare resonance detuning that is independent of R), {an additional tuning method\cite{Krenner_PRL2005,Scheibner_NPhys2008} is clearly required for a more quantitative measure of the interaction-mediated splitting}.

{To compare the experiments directly with theory is difficult because 
we do not know the nominal splittings (i.e., without dipole-dipole coupling) in the experiments, nor the dipole moments of the QD excitons. However, in an attempt to
connect the two,  Fig.~\ref{fig:Exp-v-Theory} shows the analysis of spectral splitting in the closest QD pairs (i.e. 15~sec/5~nm), where we have fit the dipole moment, $d_{\rm fit}$, given a fixed separation of 5~nm. To fit the data, we use the GF theory as presented and analyzed in Secs.~\ref{sec:theory} and \ref{sec:results}, and make the assumption that there are two excitons -- one in $\hat{\mathbf{x}}$ and one in $\hat{\mathbf{y}}$. In the case of one dipole moment, $d_{\rm fit}$ increases by a factor of $\sqrt{2}$, and the number would reduce if we allowed polarization cross coupling. Figure~\ref{fig:Exp-v-Theory} also shows the fit for different nominal detunings of 0, 5, and 10~meV, which ranges generously over the expected detuning for such QD pairs. From these fits, we see that average dipole moment required to explain the experimental results is approximately 80-100~D. Of course, the splitting is dependent on both $R$ and $d$ (approximately, $\delta\propto\frac{d^2}{R^3}$ in the near field), so slight variation in the measured separation will affect the fitted dipole moment. However, the QD separation was determined quite precisely from calibration samples, as shown by SEM imaging (Fig.~\ref{fig:SEM_TEM}).}



These extracted dipole moments are larger than what might be expected (e.g., say 30-60 D), suggesting that there are other effects going on in the experiments beyond dipole-dipole coupling (though we also include effects beyond the usual static dipole-dipole coupling term).
 However, it does not consider charge-tunnel-mediated coupling between the electronic states in each dot. Interestingly, however, both approaches can be described by a Hamiltonian of the form given in Eq.~\eqref{ham} when including the full GF response within the electric field operator, hence the predicted spectral splitting as a function of $R$ can be qualitatively similar~\cite{Bayer_Sci2001,Bester_PRL2004,Swiderski_PRB2017}
for the main spatially dependent coupling
rate. Indeed, even at the classical Maxwell level, the optical near field optical coupling is well known to reproduce the expected F\"orster coupling~\cite{Thomas2002}.
However, quantitative differences are to be expected since the dipole-dipole model does not include the excited states of the quantum dots whereas the tunneling approach neglects the possibility of long range interaction, and our approach allows one to more easily account for nonlinear and quantum optical processes.
As we have also discussed earlier, for QD disks whose radii become larger than the vertical separation, the
dipole approximation for the emitter-field
interaction is likely to breakdown.

\section{Conclusions}
\label{sec:con}

We have first presented a theoretical GF analysis of the linear spectra from two QDs (QD molecules) in a InP nanowire waveguide coupled via photon mediated dipole-dipole interactions, including effects beyond the usual static coupling. We introduce appropriate analytic solutions for the waveguide GF as well as dipole-dipole coupling via the homogeneous GF, alllowing us to model the photon transport along the wire. Using quantitative FDTD numerical calculations for the full 3D structure, we then show how both the background and waveguide contributions of the GF, as well as including retardation effects, are required to adequately model the spectral splitting of the QD resonance. Using these quasi-analytic solutions, we examine the spectral splitting and the FWHM of the dressed states of the QD molecule system as a function of homogeneous QD broadening, detuning, and spatial separation of the QDs along the axis of the nanowire. Second, we  presented a quantum master equation approach to better examine the non-linear spectra due to an increasing incoherent pumping strength, revealing a reversal in the relative spectral weight and linewidths of the peaks in the emission spectrum in the high pump regime. In the limit of weak pumping, this approach recovers the GF linear spectra within the Markov approximation. 

Next, we provided experimental PL measurements from nominally identical QDs in InP nanowires which show clearly increased splitting with decreasing spatial separation. Two types of growth are performed: diameter dependent growth and time dependent growth, the first of which has a strong shift in the resonant frequency of the QDs due to the changing diameter, and thus, changing electron-hole wavefunctions.
{We also showed a summary of the experimental
data for closely separated QDs, and extracted the
theoretically determined dipole moments to yield the same
splitting, which suggested dipole moments 
of around 80-100~D, which are likely too high (though unknown for our QD disks).}
It is likely that the dipole model is not sufficient in modelling such systems which have lateral radii on the same order of magnitude as the separations (similar effects happen for QD disks approaching metal surfaces~\cite{jun_ahn_2003}), and this could be interesting to explore in future work. {Nevertheless, there is compelling evidence that there is certainly pronounced
QD couplings, even for nominal exciton separations of
around 5~meV, and the expected dipole-dipole splittings 
can likely be considered a lower limit.} \\ 

\acknowledgments
This work was supported by the Natural Sciences and Engineering Research Council of Canada, Queen’s University, and Lumerical Solutions Inc. We  thank Mohsen Kamandar Dezfouli and
Marek Korkusinski
for useful discussions.

\appendix

\section{Beyond the dipole approximation: the Lorentz oscillator model in FDTD}
\label{app:1}

As an alternative to using a simple dipole emitter in FDTD to get the nanowire GF without embedded QDs, we may model the QDs directly as a finite-size Lorentz oscillator (LO), as outlined in Schelew \textit{et al.} \cite{Schelew2017}, and numerically obtain ${\bf G}^{(1)}$ (e.g., in the case of one QD). In that work, a single LO was implemented in Lumerical FDTD in two ways: a single-Yee cell dot and a multi-Yee cell dot (both spherical). The purpose of using such a model for dot-dot coupling is to capture any cross coupling beyond the dipole approximation due to the geometry of the QDs. The LO permittivity, $\epsilon^{\rm{LO}}$, is defined as
\begin{equation}
\epsilon^{\rm{LO}} = \epsilon_{\rm B} + \frac{\alpha_0(\omega)/V}{1 - \alpha_0(\omega)[1 - (\zeta k_{\rm B}r)^2]/(3\epsilon_{\rm B}V)}, 
\label{epsLO}
\end{equation}  
where $V$ is the volume of the disk ($V = 2\pi r^2\Delta z$ for a disk, $V=\Delta z^3$ for a single-Yee dot), $r$ is the disk radius, $\Delta z$ is the mesh size in $\hat{\mathbf{z}}$ (all directions), $\zeta$ is a correction factor (based loosely on the Mie frequency shifts for a sphere), and $\alpha_0$ is the bare polarizability of a point LO dipole, defined by Equation~\ref{alpha}. Since QDs have a geometry that is disk-like, we predict that as the distance between the dots becomes comparable to the radius of the dots, there may be enhanced coupling (and thus splitting) due to cross coupling $x-y$ terms in the GF.  
 Although these results show spurious modes due to finite gridding effects, they generally support the overall splitting estimated from the point dipole model. A more detailed analysis of finite-size dots would require coupling the QD wave functions with the GFs, which is beyond the scope of the present paper. Techniques for modelling single QD dipole breakdown effects are discussed in Refs. \onlinecite{jun_ahn_2003, stobbe_2012,tighineanu_2015}.

\begin{figure*}[th]
\centering
\subfloat{\includegraphics[trim= 0.1cm 0.25cm 1.75cm 0.2cm, clip=true,width=0.49\textwidth]{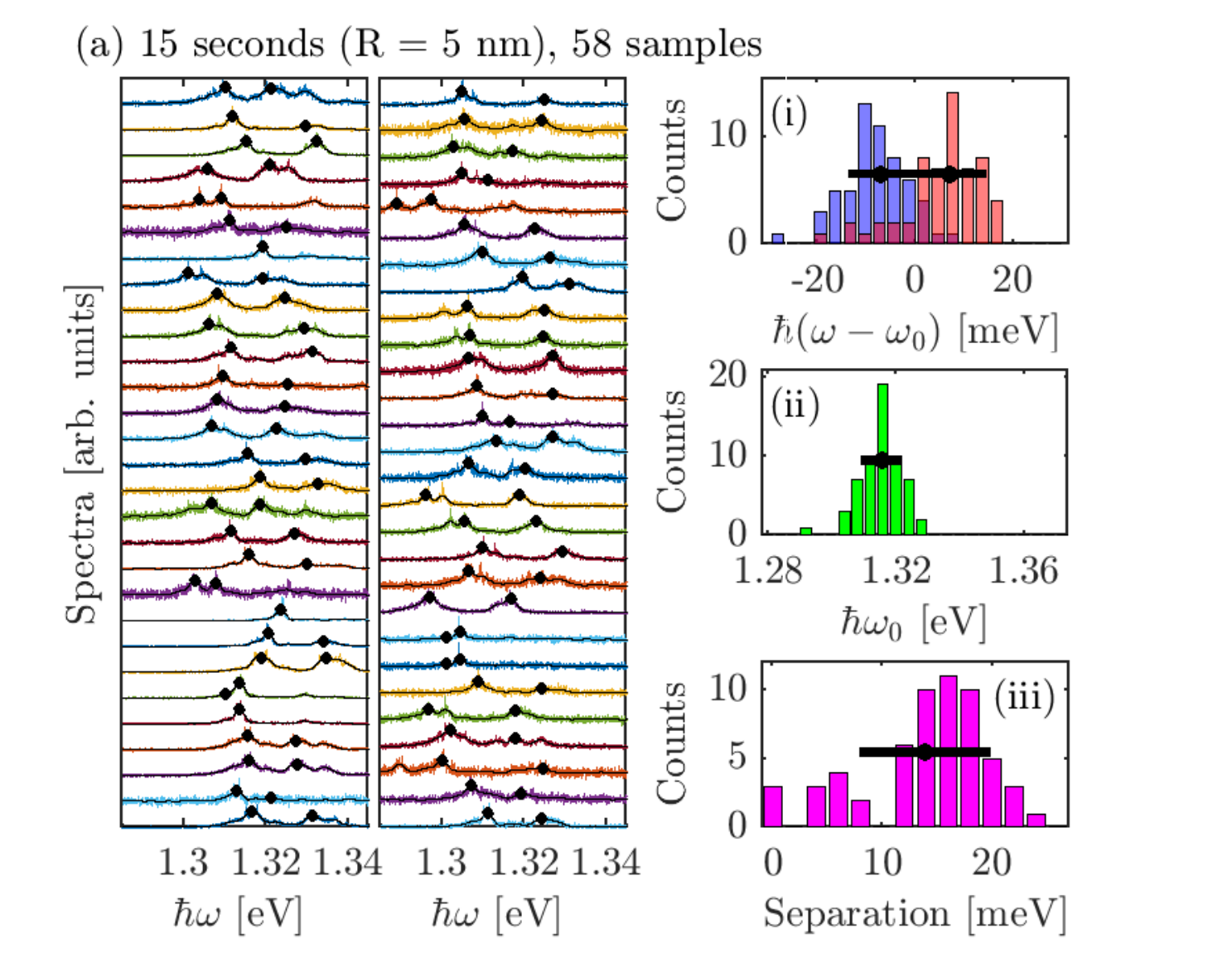}}
\subfloat{\includegraphics[trim= 0.1cm 0.25cm 1.75cm 0.2cm, clip=true,width=0.49\textwidth]{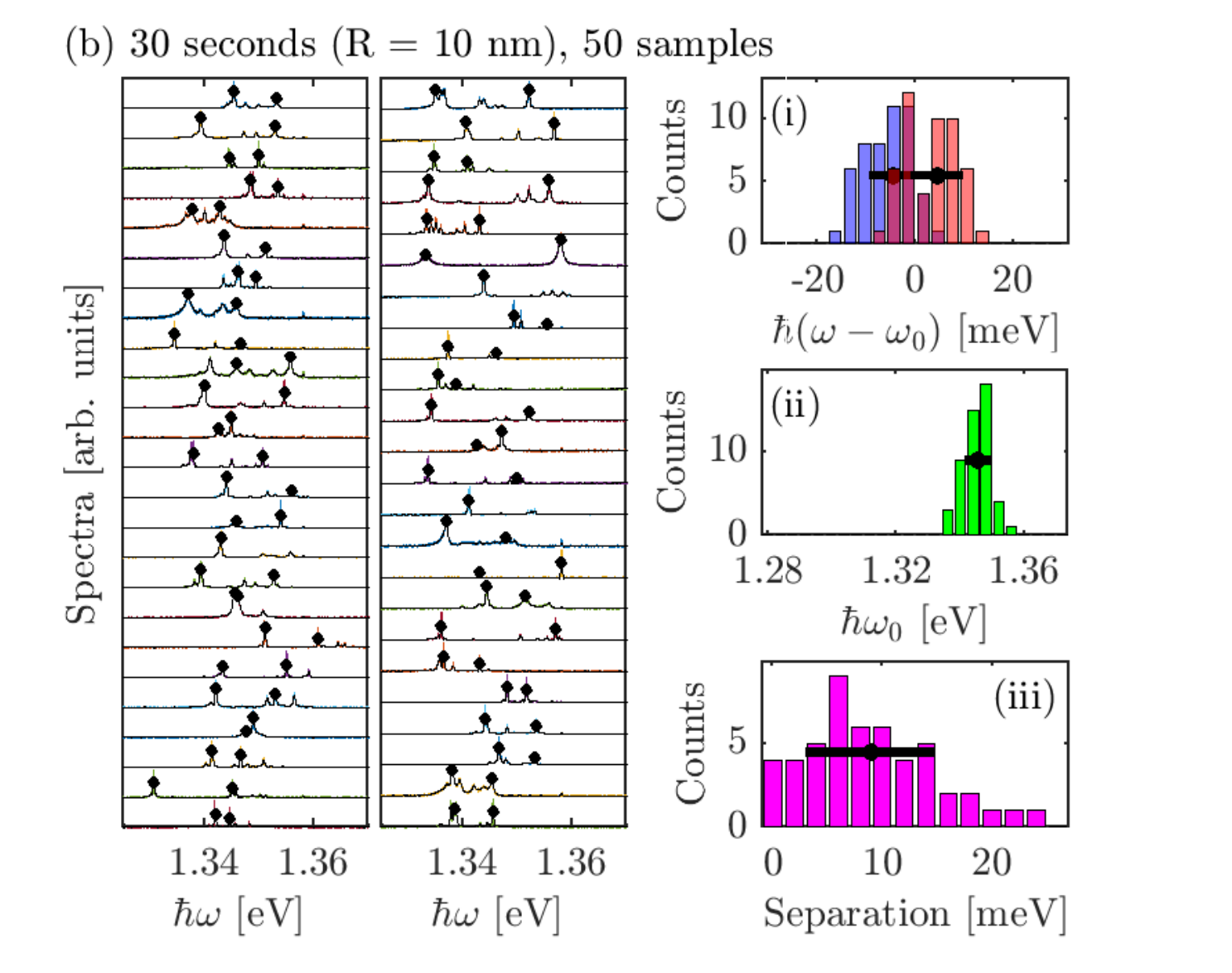}}\\
\subfloat{\includegraphics[trim= 0.1cm 0.25cm 1.75cm 0.2cm, clip=true,width=0.49\textwidth]{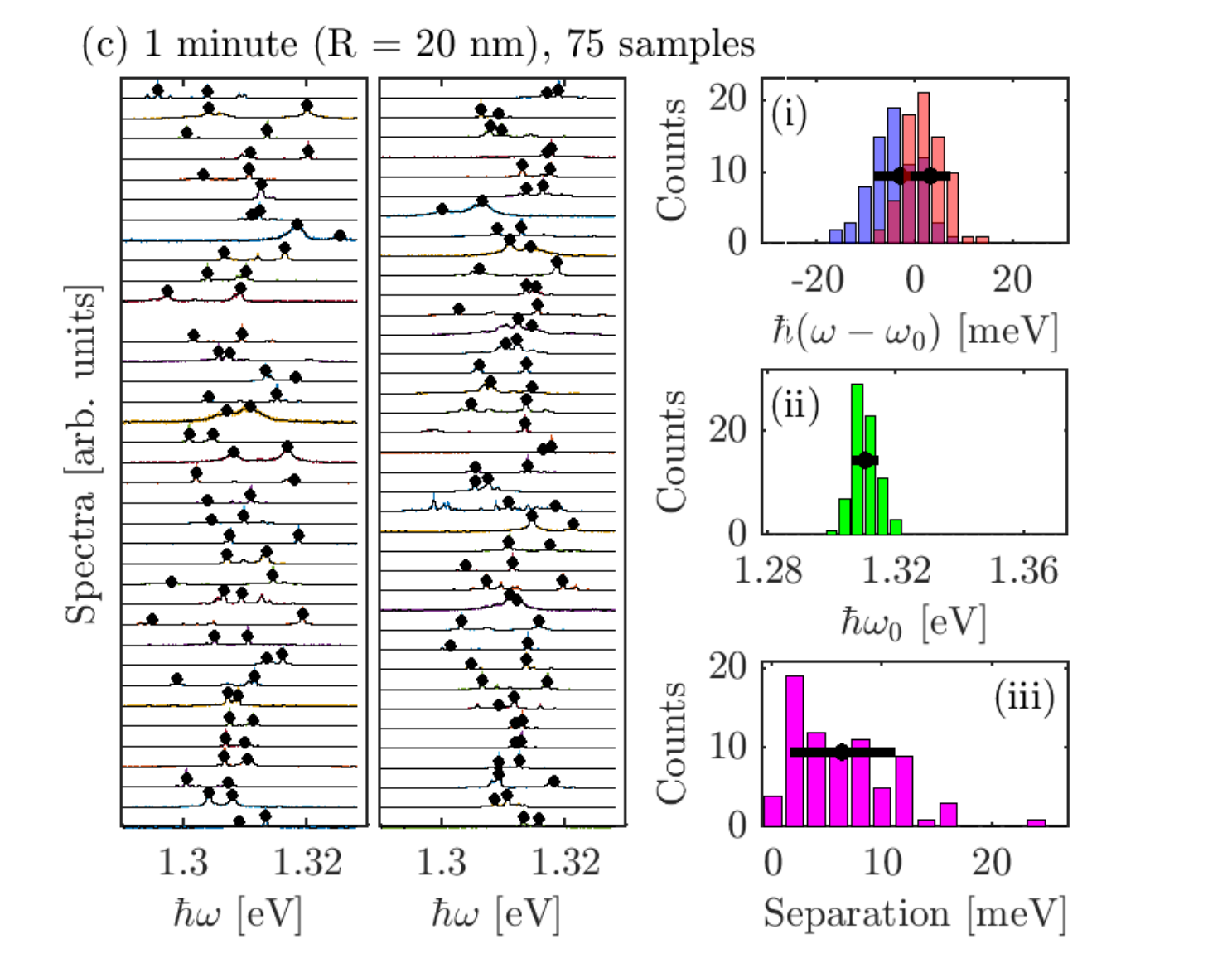}}
\subfloat{\includegraphics[trim= 0.1cm 0.25cm 1.75cm 0.2cm, clip=true,width=0.49\textwidth]{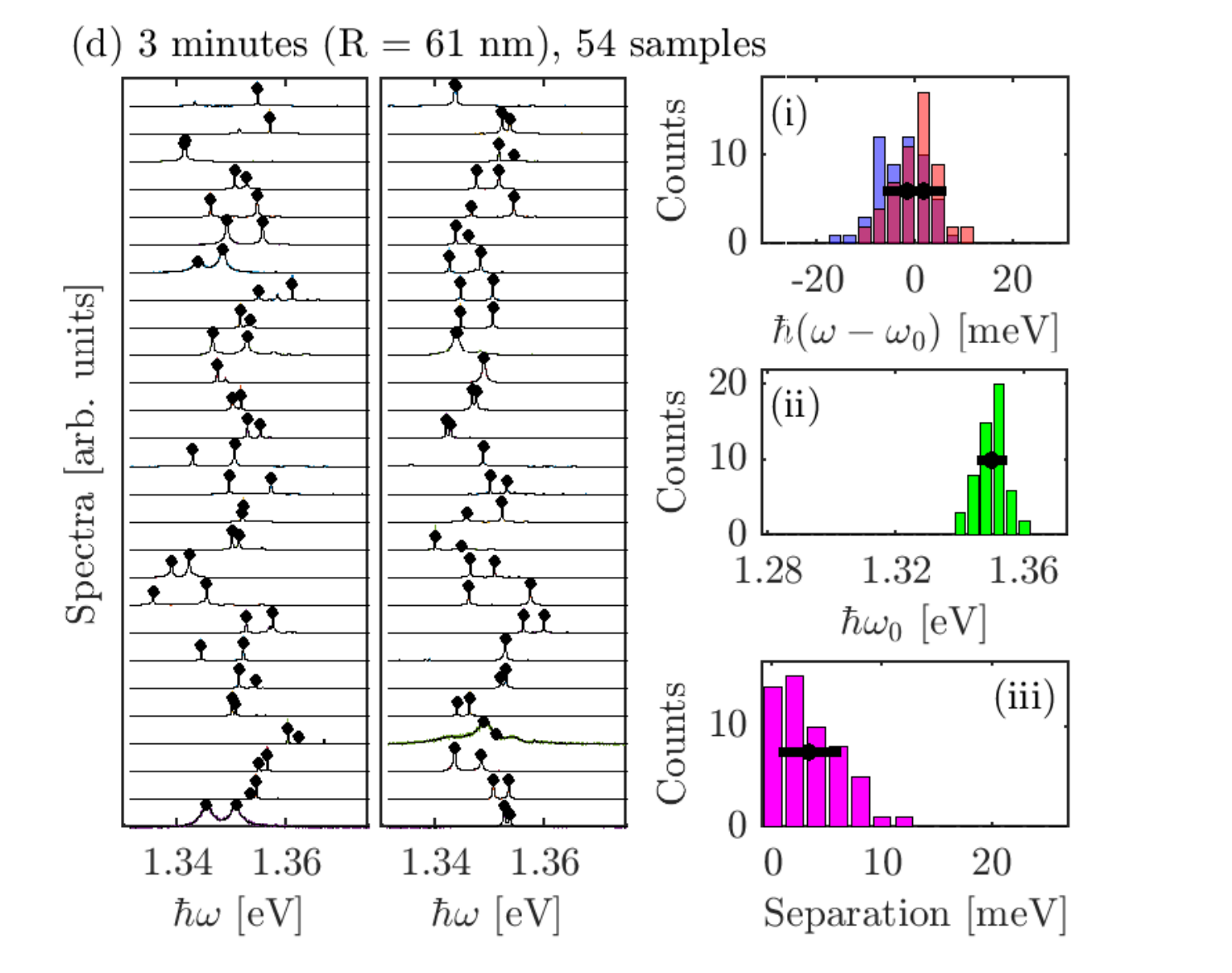}}
\caption{(a-d, left) Spectra for various sets of experimentally grown QD nanowires for different QD separations along the axis of the NW, in units of growth time (see text). {Each spectra corresponds to a specific NW grown under the required conditions for its nominal QD separation.}  Colored lines represent to the original data, and black lines are the smoothed data which reduces the noise, particularly in (a). The black markers indicate the chosen peaks used to estimate the spectral splitting due to dot-dot coupling in the NW. Excitation power is 2~nW, except for $R$=61~nm, which is 10~nW. (a-d, right) Analysis of the splitting is presented using histograms. The lower (blue) and higher (red) energy peaks are presented in (i), relative to the center energy of that dataset (ii). The total separation is shown in (iii), determined as the difference between the left and right peak in each data set (not the difference of the average left and right peaks). The histogram bin sizes (total number) are 3~meV (21), 4~meV (24), and 2~meV (14), for (i), (ii), and (iii), respectively.}
\label{fig:FullExp}
\end{figure*}

\section{Nonlinear Spectrum from Master Equation Solution}
\label{appendix:ME}
To calculate the coupled QD emitted spectrum with a master equation approach, one needs to calculate the two-time correlation functions $\langle \sigma^+_i(t+\tau)\sigma^-_j(t)\rangle$, for $i,j = 1,2$. With the aid of the quantum regression theorem and a Born-Markov approximation (also made in the derivation of the master equation), these two-time correlation functions can be found from the master equation solution:
\begin{equation}
\langle \sigma^+_i(t+\tau)\sigma^-_j(t)\rangle = \text{Tr}[\sigma_i^+\tilde{\rho}^{(j)}(\tau)],
\end{equation}
where the function $\tilde{\rho}^{(j)}(\tau)$ evolves in the same manner as the density operator in Eq.~\eqref{master_eq}, but solved with initial condition $\sigma_j^-\rho(t)$. These two-time correlation functions are most easily found if we solve for the matrix elements of $\tilde{\rho}^{(j)}(\tau)$ in the basis of the system eigenstates $\ket{E},\ket{G},\ket{\Psi_+}\ket{\Psi_-}$. In this case, we find:
\begin{align}
& \langle \sigma^+_1(t+\tau)\sigma^-_j(t)\rangle = \frac{1}{\sqrt{2}}\big[\tilde{\rho}^{(j)}_{+,E}-\tilde{\rho}^{(j)}_{-,E}+\tilde{\rho}^{(j)}_{G,+}+\tilde{\rho}^{(j)}_{G,-}\big], \\
& \langle \sigma^+_2(t+\tau)\sigma^-_j(t)\rangle = \frac{1}{\sqrt{2}}\big[\tilde{\rho}^{(j)}_{+,E}+\tilde{\rho}^{(j)}_{-,E}+\tilde{\rho}^{(j)}_{G,+}-\tilde{\rho}^{(j)}_{G,-}\big].
\end{align}
The relevant equations for the $\tilde{\rho}^{(j)}_{a,b}(\tau)$ functions can be found from Eq.~\eqref{master_eq} as $\bra{a}\frac{\rm{d}\rho}{\rm{d\tau}}\ket{b}$:
\begin{align}
& \frac{\rm{d}\tilde{\rho}^{(j)}_{G+}}{\rm{d\tau}} = -R_{G,+}\tilde{\rho}^{(j)}_{G,+} + (\Gamma_{1,1}+\Gamma_{1,2})\tilde{\rho}^{(j)}_{+,E}, \\
& \frac{\rm{d}\tilde{\rho}^{(j)}_{+E}}{\rm{d\tau}} = -R_{+,E}\tilde{\rho}^{(j)}_{+,E} + \Gamma_{\rm{inc}}\tilde{\rho}^{(j)}_{G,+},  \\
& \frac{\rm{d}\tilde{\rho}^{(j)}_{G-}}{\rm{d\tau}} = -R_{G,-}\tilde{\rho}^{(j)}_{G,-} - (\Gamma_{1,1} - \Gamma_{1,2})\tilde{\rho}^{(j)}_{-,E},  \\
& \frac{\rm{d}\tilde{\rho}^{(j)}_{-E}}{\rm{d\tau}} = -R_{-,E}\tilde{\rho}^{(j)}_{-,E} - \Gamma_{\rm{inc}}\tilde{\rho}^{(j)}_{G,-} , 
\end{align}
with $R_{a,b}$ defined in the main text. The non-zero components of the initial conditions are $\tilde{\rho}^{(1)}_{+,E}(\tau=0) = \frac{1}{\sqrt{2}}P_E$, $\tilde{\rho}^{(1)}_{-,E}(0) = \frac{-1}{\sqrt{2}}P_E$, $\tilde{\rho}^{(1)}_{G,+}(0) = \frac{1}{\sqrt{2}}P_+$, and $\tilde{\rho}^{(1)}_{G,-}(0) = \frac{1}{\sqrt{2}}P_-$ for $j=1$, and $\tilde{\rho}^{(2)}_{+,E}(0) = \frac{1}{\sqrt{2}}P_E$, $\tilde{\rho}^{(2)}_{-,E}(0) = \frac{1}{\sqrt{2}}P_E$, $\tilde{\rho}^{(2)}_{G,+}(0) = \frac{1}{\sqrt{2}}P_+$, and $\tilde{\rho}^{(2)}_{G,-}(0) = \frac{-1}{\sqrt{2}}P_-$ for $j=2$, where $P_x = \lim_{t\rightarrow \infty}\bra{x}\rho(t)\ket{x}$. Taking the Laplace transform $f(s) = \int_0^\infty dt e^{-st}f(t)$ of the above equations and letting $s =i\omega$, one arrives at the solutions given in the main text, where a factor of $1/2$ has been dropped as it can be factored out of the overall spectrum. From inspection of these solutions and the initial conditions above, it can be seen that $S_{1,1}^0(\omega) = S_{2,2}^0(\omega)$, and $S_{1,2}^0(\omega) = (S_{2,1}^0(\omega))^*$. Thus, for notationally simplicity, we drop the $j$ index and tildes and let $\rho_{a,b}(\omega) \equiv \tilde{\rho}^{(1)}_{a,b}(s = i\omega)$. To calculate the weak excitation approximation solution also used in the main text, the same procedure as above is carried out, but in a truncated basis without the $\ket{E}$ state.

\section{Additional Experimental Data}
\label{appendix:data}

{Figure~\ref{fig:FullExp} summarizes all of the experimental data for each of the four QD separations (15 seconds, 30 seconds, 1 minute, and 3 minutes), with fixed diameters (i.e. time controlled separation verses diameter controlled separation). In general, the peaks in the spectra were chosen as the two brightest peaks, since it is not feasible to fit each spectra for the excitons ((un)charged, bi-excitons, etc.) without a complete power-dependent analysis (i.e., such as Fig.~\ref{fig:NRCspectra2}). }




\end{document}